\documentclass[aps,twocolumn]{revtex4-1}
\usepackage{amsmath}
\usepackage{color}
\usepackage{xcolor}
\usepackage{subfigure}
\usepackage[outercaption,wide]{sidecap}
\usepackage{graphicx}
\usepackage{hhline}
\usepackage{tikz}
\usepackage{tikzscale}
\usepackage{pgfplots}

\pgfplotsset{compat=1.13}
\usepgfplotslibrary{fillbetween, colormaps}
\usetikzlibrary{calc,patterns,angles,quotes, arrows.meta, decorations.pathmorphing, math, decorations.pathreplacing, fadings, bending, positioning, decorations.text}

\newcommand \rmu{{{\rm u}}}

\newcommand \tT{{{\tilde{T}}}}

\newcommand \syy{{{\sigma_{yy}}}}

\newcommand \sxx{{{\sigma_{xx}}}}
\newcommand \sxy{{{\sigma_{xy}}}}
\newcommand \xmax{{{x_{max}}}}

\newcommand \Ksub{K_{\rm sub}}

\newcommand \beq{\begin{equation}}
\newcommand \eeq{\end{equation}}
\newcommand \fLW{\tfrac{L}{W}}

\begin{document}


\title{Stretching Hookean ribbons \\
Part I:
relative edge extension underlies transverse compression \& buckling instability}
%
\author{Meng Xin and Benny Davidovitch}
\affiliation{Physics Department, University of Massachusetts, Amherst MA 01003}
\begin{abstract}
The wrinkle pattern exhibited upon stretching a rectangular sheet has attracted considerable interest in the ``extreme mechanics'' community. Nevertheless, key aspects of this notable phenomenon remain elusive. Specifically -- what is the origin of the compressive stress underlying the instability of the planar state? what is the nature of the ensuing bifurcation? how does the shape evolve from a critical, near-threshold regime to a fully-developed pattern of parallel wrinkles that permeate most of the sheet? In this paper we address some of these questions through numerical simulations and analytic study of the planar state in Hooekan sheets. We show that transverse compression is a boundary effect, which originates from the relative extension of the clamped edges with respect to the transversely-contracted, compression-free bulk of the sheet, and draw analogy between this edge-induced compression and Moffatt vortices in viscous, cavity-driven flow. Next we address the instability of the planar state and show that it gives rise to a buckling pattern, localized near the clamped edges, which evolves -- upon increasing the tensile load -- to wrinkles that invade the uncompressed portion of the sheet. Crucially, we show that the key aspects of the process -- from the formation of transversely-compressed zones, to the consequent instability of the planar state and the emergence of a wrinkle pattern -- can be understood within a Hookean framework, where the only origin of nonlinear response is geometric, rather than a non-Hookean stress-strain relation.  


\end{abstract}
\maketitle

\section{Introduction}

\begin{figure*}[t]
\begin{tikzpicture}
    \node at (5.65, 2.6) {\includegraphics[width=0.33\linewidth]{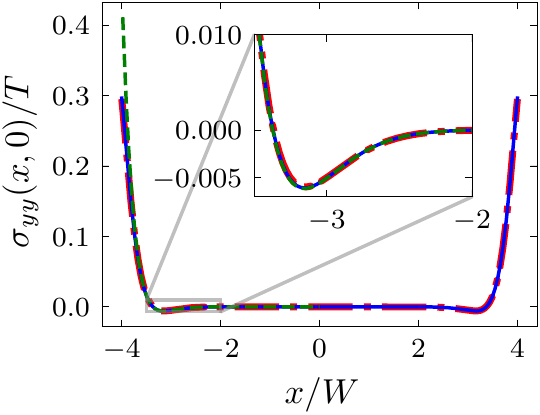}};
    \node at (5.65, -2.6) {\includegraphics[width=0.33\linewidth]{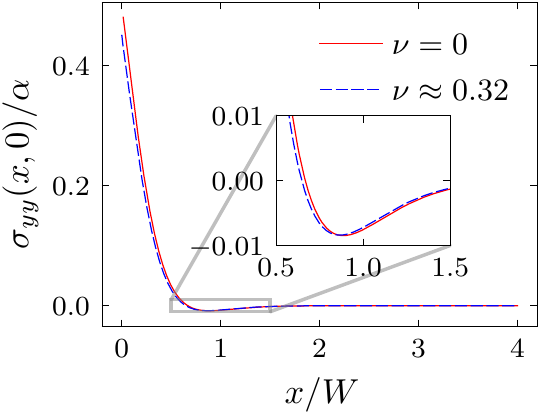}};
    \node at (2.9, 4.6) {(C)};
    \node at (2.9, -0.6) {(D)};
    \node at (-3.3, 0) {
        \begin{tikzpicture}
        \node at (-2.3, 5.2) {(A)};
        \node at (0.2, 4.4) {\resizebox{0.33\linewidth}{!}{\begin{tikzpicture}[scale=1]
\Large
    \tikzmath{\xshift=0.3; \yshift=0.2; \halfw=1; \halfl=2.5; };
\coordinate (LU) at (-\halfl, \halfw); 
\coordinate (LL) at (-\halfl, -\halfw);
\coordinate (RU) at (\halfl,\halfw);
\coordinate (RL) at (\halfl,-\halfw);
\coordinate (rLU) at ($(LU) + (-\xshift, 0) $); 
\coordinate (rLL) at ($(LL) + (-\xshift, 0) $);
\coordinate (rRU) at ($(RU) + (\xshift, 0) $);
\coordinate (rRL) at ($(RL) + (\xshift, 0) $);
\coordinate (sLU) at ($(-1, \halfw) + (0, -\yshift)$);
\coordinate (sLL) at ($(-1, -\halfw) + (0, \yshift)$);
\coordinate (sRU) at ($(1, \halfw) + (0, -\yshift)$);
\coordinate (sRL) at ($(1, -\halfw) + (0, \yshift)$);
\draw[dashed, thick] (LL) rectangle (RU);
\draw[thick] (rLU) -- (rLL);
\draw[thick] (rRU) -- (rRL);
\draw[thick] plot [smooth, tension=0.5] coordinates {(rLU) (sLU) (sRU)  (rRU)};
\draw[thick] plot [smooth, tension=0.5] coordinates {(rLL) (sLL) (sRL)  (rRL)};
\fill[pattern=north west lines] (rLL) rectangle ($(rLU) + (-0.1, 0)$);
\fill[pattern=north west lines] (rRL) rectangle ($(rRU) + (0.1, 0)$);
\draw (rLL) rectangle ($(rLU) + (-0.1, 0)$);
\draw (rRL) rectangle ($(rRU) + (0.1, 0)$);
\draw[->] ({-\halfl -\xshift -0.2}, 0) -- ++(-0.5,0) node[left] {$T$};
\draw[->] ({\halfl +\xshift +0.2}, 0) -- ++(0.5,0) node[right] {$T$};
\draw[{|Latex[width=2mm,length=2mm]}-{Latex[width=2mm,length=2mm]|}] ([yshift=-5pt]LL) -- ([yshift=-5pt]RL) node[midway, below] {$L$};
\node[right, below] at (0,0) {$O$};
\node at (0,0) {$\bullet$};
\end{tikzpicture}}};
        \node at (3.7, 5.17) {(B)};
        \node at (6,4.67) {\resizebox{0.22\linewidth}{!}{\begin{tikzpicture}[scale=1]
\Large
    \tikzmath{\stretch=0.2; \yshift=0.; \halfw=1; \halfl=2.5; };
\coordinate (LU) at (-\halfl, \halfw); 
\coordinate (LL) at (-\halfl, -\halfw);
\coordinate (RU) at (\halfl,\halfw);
\coordinate (RL) at (\halfl,-\halfw);
\coordinate (rLU) at ($(LU) + (0, \stretch) $); 
\coordinate (rLL) at ($(LL) + (0, -\stretch) $);
\coordinate (rRU) at ($(RU) + (0, \stretch) $);
\coordinate (rRL) at ($(RL) + (0, -\stretch) $);
\coordinate (sLU) at ($(-1, \halfw) + (0, -\yshift)$);
\coordinate (sLL) at ($(-1, -\halfw) + (0, \yshift)$);
\coordinate (sRU) at ($(1, \halfw) + (0, -\yshift)$);
\coordinate (sRL) at ($(1, -\halfw) + (0, \yshift)$);
\draw[dashed, thick] (LL) rectangle (RU);
\draw[thick] (rLU) -- (rLL);
\draw[thick] (rRU) -- (rRL);
\draw[thick] plot [smooth, tension=0.5] coordinates {(rLU) (sLU) (sRU)  (rRU)};
\draw[thick] plot [smooth, tension=0.5] coordinates {(rLL) (sLL) (sRL)  (rRL)};
\draw[->] ({-\halfl -0.2}, 0.2) -- ++(0, \halfw);
\draw[->] ({-\halfl -0.2}, -0.2) -- ++(0, -\halfw);
\draw[->] ({\halfl +0.2}, -0.2) -- ++(0, -\halfw);
\draw[->] ({\halfl +0.2}, 0.2) -- ++(0, \halfw);
\draw[{|Latex[width=2mm,length=2mm]}-{Latex[width=2mm,length=2mm]|}] (sLU) -- (sLL) node[midway, left] {$W$};
\node[right, below] at (0,0) {$O$};
\node at (0,0) {$\bullet$};
\draw[<->] (0, {\halfw/2}) node[right] {$y$} -- (0,0) -- ({\halfw/2}, 0) node[right] {$x$};
\end{tikzpicture}}};
        \node at (0.2,3) {\resizebox{0.3\linewidth}{!}{\input{Figures/ModelA_NT_indicate.tikz}}};
        \node at (0.2, 1) {\includegraphics[width=0.3\linewidth]{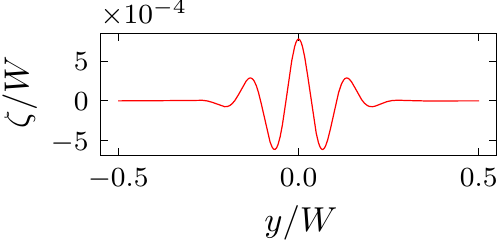}};
        \node at (0.2,-1) {\resizebox{0.3\linewidth}{!}{\input{Figures/ModelA_FFT_indicate.tikz}}};
        \node at (0.2, -3) {\includegraphics[width=0.3\linewidth]{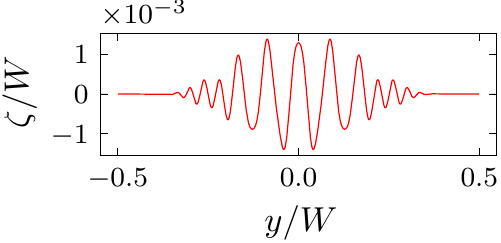}};
        \node at (6.2,3) {\resizebox{0.3\linewidth}{!}{\input{Figures/ModelB_NT_indicate.tikz}}};
        \node at(6, 1) {\includegraphics[width=0.3\linewidth]{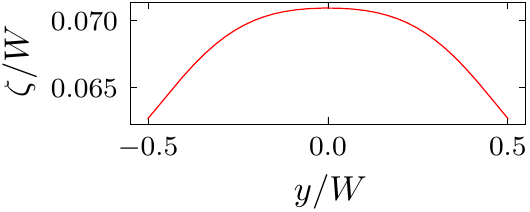}};
        \node at (6.2,-1) {\resizebox{0.3\linewidth}{!}{\input{Figures/ModelB_FFT_indicate.tikz}}};
        \node at (6, -3) {\includegraphics[width=0.3\linewidth]{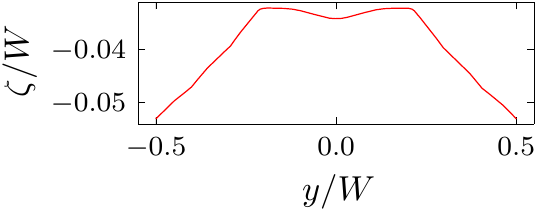}};
        \node at (-2.3, 3.2) {(E)};
        \node at (3.7, 3.2) {(F)};
        \node at (-2.3, -0.8) {(G)};
        \node at (3.7, -0.8) {(H)};
\end{tikzpicture}
    };
    
\end{tikzpicture}

\caption{
(A) Schematic drawing of ``model A'': a rectangular sheet with width $W$ and length $L$, subjected to longitudinal tensile loads, $T$ ($force/length$)  
that pull on the two short edges, $x=\pm L/2$, while the long edges are free. The short edges are clamped, such that both normal (out-of-plane) displacement, $\zeta(x=\pm L/2,y)$, and transverse (in-plane) displacement, $\rmu_y(x=\pm L/2,y)$, 
vanish, 
and their longitudinal displacements are $\rmu_x(x=\pm L/2,y) \approx \pm \tT L/2$.  
(B) Schematic drawing of ``model B'': a similar sheet is subjected to uniform stretching of its two short edges, such that $\partial_y\rmu_y(x=\pm L/2,y)  = \alpha$, while $\zeta(x=\pm L/2,y)=\rmu_x(x=\pm L/2,y)=0$.
(C) The transverse component of the planar stress, evaluated from our SE simulations along the midline, $\syy (x,y=0)$, of the stretched sheet (model A, dashed-dotted red curve) and the corners-pulled sheet (model B, with $\nu_B = 0$, solid blue curve). Also plotted is the analytic solution (dashed green). The transverse stress profile 
exhibits a non-monotonic profile, whereby each of the edges at $x = \pm \tfrac{1}{2}L$ is under transverse strain ($(1-\nu_A \tT)$ in model A, and $\alpha = (1-\nu_A \tT)$ in model B), and two corresponding transversely-compressed zones whose extent and distance from these edges $\sim W$. The level of the transverse compression is small, but nonetheless finite fraction of the transverse tension at the edge 
(the maximal compression is 
$-\syy(\pm \tfrac{1}{2}L \mp \xmax,y=0)\} \approx 0.005 T$, where $\xmax \approx 1.5 W$ ). 
{(D) A plot analogous to (C) for the corners-pulled sheet (model B) but with Poisson ratio $\nu_B =\nu_A$ (red) rather than $\nu_B = 0$ (dashed blue) exhibits an almost indistinguishable profile of the transverse stress.} 
(E-F) The deformation $\zeta(x,y)$ of the stretched sheet (model A, in panel E) and the corners-pulled sheet (model B, in panel F) in the near-threshold regime, $\tT \approx 2 \tT_c (\epsilon)$. 
For each model we show a topographic map  
and a corresponding transverse cross-section, $\zeta(x = \tfrac{1}{2}L -\xmax,y)$. 
(G-H) Deformation patterns analogous to (E-F), but in the far-from-threshold regime, $\tT \approx 118 \tT_c (\epsilon)$.   }
\label{fig:schematic}
\end{figure*}

A familiar, 
yet quite 
nontrivial pattern formation phenomenon, 
is the parallel array of wrinkles that extend throughout a ribbon -- a thin, rectangular-shaped solid sheet -- 
upon pulling its clamped edges apart 
(Fig.~\ref{fig:schematic}A) \cite{Friedl2000, Cerda-Mahadevan2002}.
Recalling that buckling and wrinkling of a thin solid body emerge in response to compression, 
one may readily conclude that the 
visible array of 
wrinkles parallel to the stretching axis, $\hat{x}$, 
is due to a compressive component of the stress tensor, along the transverse axis, $\hat{y}$. 
However, while a {\emph{transverse contraction}} ({\emph{i.e.}} negative strain, $\varepsilon_{yy} <0$) 
in response to longitudinal tension, $\sxx \approx T >0$, is the essence of the classical Poisson effect, the appearance of 
{\emph{transverse compression}} 
({\emph{i.e.}} $\syy <0$) is far less obvious.  
Indeed, if the pulled edges 
were not clamped, the whole sheet would have contracted uniformly in the transverse direction, the stress would have been perfectly uniaxial and tensile everywhere ({\emph{i.e.}} $\varepsilon_{yy} \propto -\nu T$, where $\nu$ is 
the Poisson ratio, 
and $\sxx = T \ , \ \syy = \sigma_{xy}=0)$, and the planar, unwrinkled state, would have been stable. Hence, 
the emergence of transverse compression 
is necessarily a boundary effect, which may exist only near the clamped edges. 
While numerical and analytical studies of the planar (unwrinkled) state did 
reveal 
the presence of zones with small transverse compression 
close to the clamped edges \cite{Friedl2000,Nayya2011}, 
the physical mechanism underlying this boundary effect 
remains elusive. 

Even more puzzling than the mere existence of transversely compressed zones in the planar state is the ensuing elastic instability.
Rather than forming a buckling pattern, characterized by a thickness-independent topography that   
relieves the transverse compression, 
the stretched sheet appears to develop a highly corrugated topography, 
whereby the characteristic wavelength $\lambda$ of transverse undulations has been reported to be proportional to the square root of the sheet's thickness $t$ \cite{Cerda-Mahadevan2002}. 
Elastic instabilities of thin bodies that give rise to a thickness-dependent wavelength, $\lambda \sim t^{\alpha}$ (with $\alpha >0$) are often called {\emph{``wrinkling''}}  and are known to occur in supported sheets subjected to uniaxial compression, whereby the resistance of the attached subphase to deformation competes with the tendency of the sheet to minimize bending energy 
\footnote{ ``buckling'' is thus understood as a particular instance of ``wrinkling``, where the power $\alpha=0$}. For instance, the undulation wavelength of uniaxially-compressed solid sheets that are floating on a liquid bath or attached to a compliant elastic medium scale as $\lambda \sim t$, and $\lambda \sim t^{3/4}$, respectively \cite{Milner89,Bowden98}. 
However, why does the transverse compression of a suspended sheet give rise to a highly-curved wrinkle pattern, $\lambda \sim t^{1/2}$, 
whose 
bending energy is substantially larger than a buckling pattern ($\lambda \sim t^0$) that is also capable of relieving transverse compression? 

Realizing that the observed wrinkle pattern in this system  
cannot be described  
through a standard ``post-buckling'' approach, in which the out-of-plane deflection of a (naturally planar) sheet 
is assumed to affect 
only slightly 
the planar stress, Cerda   \&   Mahadevan (CM) proposed 
to address the system in a strictly distinct,  
{\emph{``far from threshold"}} (FT) regime, $T \gg T_c$, where $T_c$ is a threshold value, below which the compressed planar state is stable
\cite{Cerda03}. 
In this approach, which is based on ``tension field theory''
\cite{Wagner29, stein61, MansfieldBook, Pipkin86, Steigman90},  
the wrinkle pattern is assumed to be fully developed throughout the whole sheet, and   
cannot be described as a perturbation to the compressed planar stress, but rather to a compression-free stress field, attained by a hypothetic sheet, with finite stretching modulus 
and no bending modulus. 
The CM model inspired a multitude of experimental and theoretical works that addressed the far-from-threshold regime of wrinkle patterns in various systems of ultarthin sheets subjected to confinement by capillary effects or other forces \cite{Paulsen19,Vella19}. 
%
%
However, the ingenuous proposal to focus on the FT parameter regime evades 
some of the basic puzzles exhibited by the system (Fig.~\ref{fig:schematic}A) that was the original object of the CM model.  
Specifically -- what is the origin of transversely compressed zones underlying the instability of the planar state? why the out-of-plane deflection of the sheet does not remain localized to these near-edge zones but develops instead into a pattern of small wrinkles that permeate the uncompressed bulk ? 

In this paper and a subsequent one we revisit the uniaxial stretching of a Hookean ribbon-shaped sheet with clamped edges, and address the planar state, its instability, and the transition from the NT regime, $T \gtrsim T_c$, to the far from threshold regime, $ T \gg T_c$. A primary tool that we employ in our studies 
is numerical simulations with Surface Evolver (SE) \cite{Brakke1992}, 
which we find to be an excellent method for finding the planar state as well as the energetic minimum of fully developed wrinkled states in the far-from-threshold regime of very thin sheets.  
Our rationale in focusing on Hookean elasticity ({\emph{i.e.}} a linear stress-strain relationship) is twofold.  First, despite the practical importance of 
effects associated with non-Hookean response, most notably the re-stabilization of a planar, unwrinkled state when the imposed tensile strain is sufficiently large ($0.3-0.5$ \cite{Nayya11, Healy13, Healy16, Sipos16, Fu19, Wang19, Panaitescu2019}), the Hookean response unravels the universal ({\emph{i.e.}} material-independent) mechanism through which transverse compression emerges and wrinkles develop. Second, since the instability threshold $T_c$ vanishes rapidly with the sheet's thickness (more precisely, $T_c$ vanishes with thickness $t$ faster than the stretching modulus), Hookean mechanics is expected to govern for sufficiently thin sheets      
not only the transversely-compressed planar state and its instability, but also the fully developed wrinkle pattern in the       
far-from-threshold regime, $T \gg T_c$. 
%
The focus of this paper is the planar stress and the characterization of its instability. In the subsequent article we address the evolution of this instability from the near-threshold regime to a fully developed wrinkle pattern in the far-from-threshold regime.  \\  

Starting in Sec.~\ref{sec:planar-state} with numerical and analytical study of the 
planar state, we show that the ultimate cause of transverse compression is the 
extension of the clamped edges 
relative to the transversely contracted bulk of the sheet. We elucidate this subtlety by analyzing a specific set-up 
(model B in Fig.~\ref{fig:schematic}) and demonstrate how transverse compression in a rectangular sheet with free long edges can occur 
even without exerting longitudinal tensile load.     
In Sec.~\ref{sec:buckling} we address the instability of the planar stress  
and show that it is essentially an 
Euler buckling, whose spatial extent is restricted to the transversely-compressed zones of the sheet. A direct corollary of this observation   
is that the emergence of wrinkles 
in this set-up, whereby the wavelength vanishes with the thickness of the sheet, does not occur at the near-threshold regime   
($T \gtrsim T_c$); instead, it may 
only be observed in the far-from-threshold regime, $T \gg T_c$. In Sec.~\ref{sec:transition} we provide evidence for the transition from near-threshold buckling to far-from-threshold wrinkling, and defer an in-depth study of the latter to a subsequent paper. 






\section{The planar state}
\label{sec:planar-state}
The observation that longitudinal tension {\emph{does not}} induce transverse compression if the short edges are unclamped (or alternatively if $\nu \leq 0$) suggests that the primal cause for transverse compression is neither uniaxial tension nor positive Poisson ratio, but rather a relative extension of the short edges in comparison to the bulk of the sheet. 
In order to elucidate this geometrical-mechanical effect, we contrast in this section the planar state of our set-up, hence called ``model A'' (Fig.~\ref{fig:schematic}A), with the planar state of another system, called ``model B''  (Fig.~\ref{fig:schematic}B), in which a relative extension of the edge is imposed directly on a rectangular sheet, with arbitrary Poisson ratio  
and no longitudinal tension.

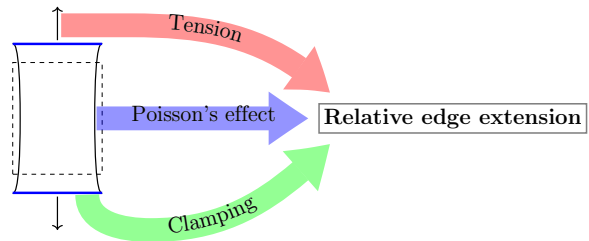
\begin{figure}
\centering
\resizebox{0.9\linewidth}{!}{\tikzset{
    vecArrow/.style args={#1, color=#2, line width=#3, raise=#4}{
      draw, #2, line width=#3, -{.Triangle[bend,angle=60:1pt 2]}, shorten >=6pt, postaction=decorate, decoration={text along path, text color=black!100, raise=#4, text align=center, text={#1}}
    }
} 
\begin{tikzpicture}[scale=0.85,
    ]
    \draw[dashed] (-1.2,-1.5) rectangle (1.2,1.5);
    \draw[thick] plot[smooth, tension=2] coordinates {(-1.2,2) (-1,0) (-1.2,-2) };
    \draw[thick] plot[smooth, tension=2] coordinates {(1.2,-2) (1,0) (1.2, 2)};
    \draw[ultra thick,blue] (-1.2,2) -- (1.2,2);
    \draw[ultra thick,blue] (-1.2,-2) -- (1.2,-2);
    \draw[->, thick] (0,2.1) -- (0,3);
    \draw[->, thick] (0,-2.1) -- (0,-3);

    \begin{scope}[opacity=.42, transparency group]
      \path [vecArrow={{|\Large|Tension ||}, color=red, line width=4ex, raise=-0.5ex}] (.1,2.5) to[in=180, out=0] (2, 2.5) to[out=0, in=135] (7.5,0.5);
      \path [vecArrow={|\Large|Poisson's effect ||, color=blue, line width=4ex,  raise=-0.5ex}] (1.05,0) to[out=0, in=180] (7,0);
      \path [vecArrow={|\Large|Clamping ||, color=green, line width=4ex,  raise=-0.5ex}] (0.8, -2.05) to [out=-90, in=180] (2.5,-3) to [out=0, in=-135]  (7.5,-0.5);
    \end{scope}

    \node[right, line width=1pt, draw=gray] at (7,0) {\bf\Large Relative edge extension};
    \draw[line width=0ex, postaction=decorate, decoration={text along path, text color=black, text align=center, text={|\Large|Clamping ||}, raise=-0.5ex}, opacity=0, fill opacity=1] (0.8, -2.05) to [out=-90, in=180] (2.5,-3) to [out=0, in=-135]  (7.5,-0.5);
    \draw[line width=0ex, postaction=decorate, decoration={text along path, text color=black, text align=center, text={|\Large|Poisson's effect ||}, raise=-0.5ex}, opacity=0, fill opacity=1] (1.05,0) to[out=0, in=180] (7,0);
     \draw[line width=0ex, postaction=decorate, decoration={text along path, text color=black, text align=center, text={|\Large|Tension ||}, raise=-0.5ex}, opacity=0, fill opacity=1] (.1,2.5) to[in=180, out=0] (2, 2.5) to[out=0, in=135] (7.5,0.5);
 \end{tikzpicture}}
\caption{
The identical transverse stress profile in model A and model B indicates that the ultimate cause for a 
transversely compressed zone in the planar state 
is a relative extension of the short edge with respect to the bulk. This may be a direct outcome of pulling the short edge outward (model B) 
or the collective effect of imposing longitudinal tension, clamping the short edge, and positive Poisson ratio (model A).
}
 \label{fig:mapping}
\end{figure}



\subsection{Displacement, strain, and Hookean mechanics}
Since our focus here is on small strains 
we express the components of the strain tensor $u_{ij}$ 
through derivatives of the displacement field, whose in-plane components are $u_x,u_y$ and whose out-of-plane component is $\zeta$: 
\begin{gather}
    u_{xx} = \partial_x u_x + \tfrac{1}{2}(\partial_x\zeta)^2 \ ; \ 
    u_{yy} = \partial_y u_y + \tfrac{1}{2}(\partial_y\zeta)^2 \nonumber   \ ; \\
    u_{xy} = \tfrac{1}{2}(\partial_x u_y + \partial_y u_x + \partial_y\zeta \partial_x\zeta)  
    \label{eq:strain-disp} 
    \ , 
\end{gather}
and invoke the Hookean stress-strain relationship \cite{LandauTE}: 
\begin{gather}
\sxx = \tfrac{1}{1-\nu^2}Y (u_{xx} + \nu u_{yy}) \  ; \syy = \tfrac{1}{1-\nu^2} Y (u_{yy} + \nu u_{xx}) \ ; \nonumber \\ \sxy = \tfrac{1}{1+\nu}Y u_{xy}  \ . 
\label{eq:stress-strain}
\end{gather}
The two problems we address in this section, ``model A'' and ``model B'', are defined below through suitable boundary conditions (BCs) on the stress tensor and the displacement field. Since we consider only small strains, 
we employ the standard approach of Hookean elasticity theory, and assume the BCs hold at the edges of the original, undeformed sheet, namely, the long edges $y = \pm \tfrac{1}{2} W $ and the short edges $x = \pm \tfrac{1}{2}L $. 
\footnote{The error incurred by considering BCs through the undeformed, rather the deformed sheet, is a higher order in $T/Y$.}

In analyzing the planar state we consider a displacement field with $\zeta=0$. A useful tool for this analysis is the Airy potential -- a scalar function $\Phi(x,y)$ such that: 
\begin{gather}
\syy = \frac{\partial^2 \Phi}{\partial x^2} \ ; \ \sxx = \frac{\partial^2 \Phi}{\partial y^2} \ ; \ 
\sxy = - \frac{\partial^2 \Phi}{\partial x\partial y}  \ . \label{eq:airy-stress} 
\end{gather} 
The mechanical equilibrium equation, $\partial_j\sigma_{ij} = 0$, becomes the bi-harmonic equation for the Airy potential: 
\begin{gather}
    \label{eq:ariy}
    \nabla^4 \Phi = 0 \ . 
\end{gather}

\subsection{model A {\emph{versus}} model B} 
The mathematical description of our original set-up (``model A'', Fig.~\ref{fig:schematic}A) consists of a non-homogeneous BC: 
\begin{subequations}\label{eq:bounday-model-A}
\begin{equation}
\int_{-\frac{W}{2}}^{\frac{W}{2}} \sxx(x=\pm \tfrac{1}{2}L,y) dy = 
T W  \ , 
\label{eq:nonhom-1}
\end{equation}
expressing the fact that a force $TW$ is pulling each of the short edges outward (and applies also for any $-\tfrac{1}{2}L\leq x \leq \tfrac{1}{2}L$ by force balance consideration). Additionally, there are
four homogeneous BCs:
\begin{align}
\text{at} & \ y=\pm \tfrac{1}{2}W:  \ \    \syy = \sxy = 0      \label{eq:b2} \\ 
\text{at} & \ x=\pm \tfrac{1}{2}L:  \ \   
u_y = 0 \ \ ; \ \ 
 \frac{\partial u_x}{\partial y} = 0  \ .    \label{eq:b3}
\end{align}
The first two equations~(\ref{eq:b2}) reflect the fact that the long edges are free, namely, $ \sigma_{ij}n_j = 0$, 
where $\hat{n}=\pm \hat{y}$ is the outward normal to the long (undeformed) edges, respectively. The last two equations~(\ref{eq:b2}) imply that the short edges are displaced as rigid, inextensible sticks, pulled apart along the $\hat{x}$ axis, such that their displacement is given by $u_x = Const$ and $u_y = 0$. 

With the aid of Eq.~(\ref{eq:airy-stress}) the BCs~(\ref{eq:nonhom-1},\ref{eq:b2},\ref{eq:b3}) may be converted to a set of four BCs for 
the Airy potential: 
\begin{align}
    \label{eq:ariybc1}
\text{at} \ y=\pm \tfrac{1}{2}W &:   \ \ \ \   \Phi = 0 \\
    \label{eq:ariybc2}
    \text{and} & \ \ \ \ \ \ 
   \frac{\partial \Phi}{\partial y} =  \pm \tfrac{1}{2}TW \\ 
    \label{eq:ariybc3}
\text{at} \ x=\pm \tfrac{1}{2}L &:   \ \ \   
\frac{\partial^2 \Phi}{\partial x^2} - \nu \frac{\partial^2 \Phi}{\partial y^2} = 0 \\
    \label{eq:ariybc4}
    \text{and} & \ \ \ \    \ \ \frac{\partial^3 \Phi}{\partial x^3}  \!+\!(2\!+\!\nu) \frac{\partial^3 \Phi}{\partial x \partial y^2} = 0 \ . 
       \end{align}
\end{subequations}
Equation~(\ref{eq:ariybc1}) follows from the first part of (\ref{eq:b2}), which implies that $\Phi(x,y = \pm \tfrac{1}{2}W) =C_0 + C_1 x$, with arbitrary constants $C_0,C_1$, upon choosing the natural gauge: $C_0 \!= \!C_1\!=\!0$. Integrating $\sxy (x,y = \pm \tfrac{1}{2}W)$ over $x$ one readily obtains from the second part of Eq.~(\ref{eq:b2}) that $\partial_y \Phi(x,\pm \tfrac{1}{2}W)$ is independent on $x$, and Eq.~(\ref{eq:ariybc2}) is thus obtained directly from Eq.~(\ref{eq:nonhom-1}), which is valid -- as explained above -- for any $-\tfrac{1}{2}L\leq x \leq \tfrac{1}{2}L$. Using the strain-displacement relations (\ref{eq:strain-disp}), Eq.~(\ref{eq:ariybc3}) follows directly from the first part of Eq.~(\ref{eq:b3}), whereas Eq.~(\ref{eq:ariybc4}) follows from the second part of~(\ref{eq:b3}) after some tedious, but straightforward algebraic manipulations.   \\

Turning now to ``model B'' (Fig.~\ref{fig:schematic}B), we describe the BCs in analogous manner to Eqs.~(\ref{eq:bounday-model-A}). First, the absence of exerted tension along the $\hat{x}$-axis implies that: 
\begin{subequations}\label{eq:bounday-model-B}
\begin{equation}
\int_{-\frac{W}{2}}^{\frac{W}{2}} \sxx(x=\pm \tfrac{1}{2}L,y) dy = 0  \ , 
\label{eq:nonhom-1B}
\end{equation}
(as well as for any $-\tfrac{1}{2}L\leq x \leq \tfrac{1}{2}L$). Second, three BCs are identical to their homogeneous counterparts in model A: 
\begin{align}
\text{at} & \ y=\pm \tfrac{1}{2}W:  \ \    \syy = \sxy = 0      \label{eq:bB2} \\ 
\text{at} & \ x=\pm \tfrac{1}{2}L:  \ \      
 \frac{\partial u_x}{\partial y} = 0 \ ,    \label{eq:bB3}
\end{align}
whereas the last BC is non-homogenous: 
\begin{equation} \label{eq:bB4}
\text{at}  \ x=\pm \tfrac{1}{2}L:  \ \      
 \frac{\partial u_y}{\partial y} = \alpha
\end{equation} 
such that $\alpha>0$ is a transverse strain imposed directly at the short edges.  
The conversion of the BCs~(\ref{eq:nonhom-1B}-\ref{eq:bB4}) into BCs for the Airy potential proceeds along the same steps that led from Eqs.~(\ref{eq:nonhom-1}-\ref{eq:b3}) to Eqs.~(\ref{eq:ariybc1}-\ref{eq:ariybc4}), yielding: 
\begin{align}
    \label{eq:ariybc1B}
\text{at} \ y=\pm \tfrac{1}{2}W &:   \ \ \ \     \Phi = 0 \\
    \label{eq:ariybc2B}
\text{and}   &   \  \ \ \ \ \ \frac{\partial \Phi}{\partial y} = 0 \\
    \label{eq:ariybc3B}
\text{at} \ x=\pm \tfrac{1}{2}L &:   \ \ \       
\frac{\partial^2 \Phi}{\partial x^2}- \nu \frac{\partial^2 \Phi}{\partial y^2}= \alpha Y \\
    \label{eq:ariybc4B}
\text{and}   &   \ \  \ \ \ \   \frac{\partial^3 \Phi}{\partial x^3} \!+\!(2\!+\!\nu) \frac{\partial^3 \Phi}{\partial x \partial y^2} = 0  . 
\end{align}
\end{subequations}
Note that the planar stress of a Hookean sheet is determined by a purely linear problem, namely, both the strain-displacement relation (\ref{eq:strain-disp}) and the stress-strain relation (\ref{eq:stress-strain}) are given by linear equations, and consequently the stress, strain, and displacement fields for each of the two models are fully determined by solving a linear PDE (\ref{eq:ariy}) subjected to the corresponding BCs (Eqs.~(\ref{eq:nonhom-1}-\ref{eq:ariybc4}) for model A or Eqs.~(\ref{eq:nonhom-1B}-\ref{eq:ariybc4B}) for model B). Furthermore, each of the two models consists of a single non-homogeneous BC (Eq.(\ref{eq:nonhom-1}) for model A and Eq.~(\ref{eq:bB4}) for model B), and therefore the stress field in each model is unique up to a scale factor (and similarly the strain and displacement fields). Namely, denoting the planar stress field for model A under a given exerted longitudinal tension $T$ by $\sigma_{ij}(x,y;T)$, and the planar stress field for model B under a given edge extension $\alpha$ by $\sigma_{ij}(x,y;\alpha)$, we have that: 
\begin{gather}
 \text{model \ A}: \ \ \sigma_{ij}(x,y;T_2) = \tfrac{T_2}{T_1} \sigma_{ij}(x,y;T_1) \nonumber \\
  \text{model \ B}: \ \ \sigma_{ij}(x,y;\alpha_2) = \tfrac{\alpha_2}{\alpha_1} \sigma_{ij}(x,y;\alpha_1)
\label{eq:lin-inv}
\end{gather}


\subsection{Numerical simulations} 
We employ SE to study the planar state of the two models, implementing an 
equilateral-triangular mesh of density $6.95\times 10^{5}$
($total\ area / cell\ area$). 
The SE built-in {\emph{``linear\_elastic"}} method is adapted for computing 
the strain energy, and {\emph{``star\_perp\_sq\_mean\_curvature"}} and {\emph{``star\_gauss\_curvature"}} to compute bending energy. 


For model A, we consider a sheet with a relatively large length-to-width ratio, $\fLW= 8$, a Poisson ratio $\nu_A = 0.32$, and some exerted longitudinal tension $T$ whose actual numerical value is arbitrary (see Eq.~(\ref{eq:lin-inv})). For model B, we consider a sheet with the same length-to-width ratio, $\fLW = 8$, and Poisson ratio $\nu_B = \nu_A$ or $\nu_B = 0$.  
Since in model B $u_{yy} = 0$ in the bulk, we   
make the extension of the short edge relative to the bulk identical to model A 
(where $u_{yy} = 0$ at the clamped edge and $-\nu_A T/Y$ in the bulk), by choosing the edge extension parameter in model B to be $\alpha = \nu_A T/Y$. 

While the longitudinal stress components, $\sxx(x,y)$, of the two models are obviously distinct, Figs.~\ref{fig:schematic}c,d show that the transverse stress, $\syy(x,y)$ in the two models is essentially identical. Furthermore, the direct effect of the Poisson ratio is negligible, as can be seen by comparing the transverse stress of model B with $\nu_B=\nu_A$ and $\nu_B=0$. As Figs.~\ref{fig:schematic}c,d show, the transverse stress is positive (tensile) in the vicinity of the short edges, becoming 
compressive at a distance  
$\approx 1.5 \cdot W$ from each short edge, and remains compressive over a strip of length $\sim  W$, after which it vanishes 
exponentially. 

Our numerical solution of the planar stress in the two models indicates that the essential cause of transverse compression in a rectangular sheet is the extension of the short edge relative to the bulk. As the schematic in Fig.~\ref{fig:mapping} shows, this effect can be attained directly (as is the case in model B) even for a sheet with $\nu=0$ with no longitudinal tensile load, or indirectly -- as in our original model A -- by applying longitudinal tension and clamping the short edges of a sheet with positive Poisson ratio.

\begin{figure}
\centering
\includegraphics[]{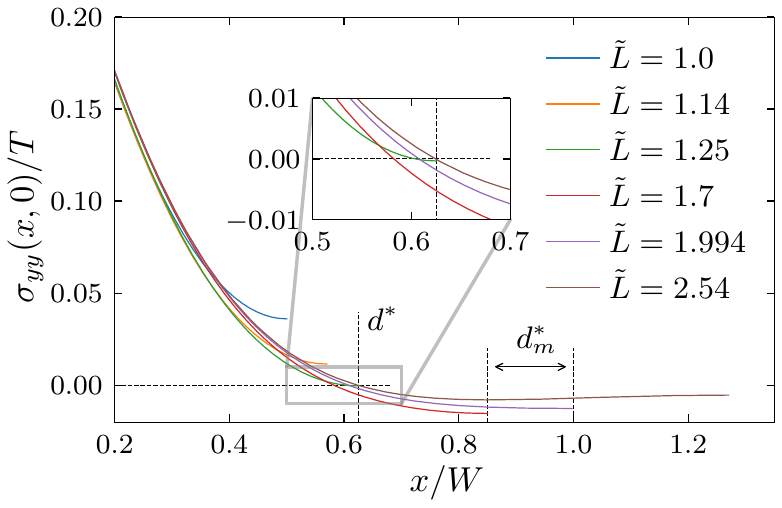}
\caption{Our SE simulations of the planar state in Hookean sheets with various aspect ratios, $L/W$, suggest the existence of a parameter regime {\emph{(i)}} where the stress may be purely tensile ans the planar state is thus stable ($\tfrac{L}{W} < 2d^*$); {\emph{(ii)}} where transverse compression exists around the middle of the sheet ($2d^*  < \tfrac{L}{W} < 2d_m^*$); and {\emph{(iii)}} where transverse compression exists in two zones 
($ \tfrac{L}{W} > 2d_m^*$). For $d^*$, the numerical values extracted from our SE simulations is $d^* \approx 0.625$, whereas for $d_m^*$ the extracted value is $d_m^* \approx 0.8 \ -- \  1.0$ (the uncertainty is a consequence of a very shallow local maximum of $\syy(x,0)$.) }  
%
 \label{fig:aspect-ratio}
\end{figure}



\subsection{Analytical solution} 
\subsubsection{Equivalence of model A and model B}
The identity of the tranverse stress components in models A and B can be understood by decomposing the Airy potential of model A into ``bulk'' and ``edge'' terms:
\begin{subequations}
\begin{gather}
\text{model \ A}: \ \Phi (x,y) = \Phi_b (x,y) + \Phi_e (x,y) \nonumber \\
\text{where}: \ \Phi_b (x,y) = \tfrac{1}{2}T (y^2-1) \ . 
\label{eq:decompose-A}
\end{gather}
If the short edges were not clamped ({\emph{i.e.}} if the first part of Eq.~(\ref{eq:b3}) {had been replaced by $\partial^2_y u_y = 0$,} such that $u_{yy}(x=\pm\tfrac{1}{2}L,y)$ may be nonzero), then $\Phi_e= 0$, and the resulting stress, associated only with $\Phi_b$, would have been constant: $\sxx = T, \syy=\sxy=0$. However, clamping implies that $\Phi_e (x,y) \neq 0 $, since it must satisfy the nonhomogenous set of BCs: 
\begin{align}
    \label{eq:ariybc1e}
    \text{at} \ y=\pm \tfrac{1}{2}W &:   \ \ \ \  
    \Phi_e = 0 \\
    \label{eq:ariybc2e}
\text{and} & \ \ \ \ \ \      \frac{\partial \Phi_e}{\partial y}  = 0 \\
    \label{eq:ariybc3e}
     \text{at} \ x=\pm \tfrac{1}{2}L &:   \ \  \    \frac{\partial^2 \Phi_e}{\partial x^2}- \nu_A \frac{\partial^2 \Phi_e}{\partial y^2} = \nu_A T \\
    \label{eq:ariybc4e}
\text{and} & \ \ \ \ \  \    \frac{\partial^3 \Phi_e}{\partial x^3} \!+\!(2\!+\!\nu_A) \frac{\partial^3 \Phi_e}{\partial x \partial y^2} = 0  . 
\end{align}
\end{subequations}
Remarkably, the BCs~(\ref{eq:ariybc1e}-\ref{eq:ariybc4e}) are identical to the BCs satisfied by the Airy potential $\Phi(x,y)$ of model B (\ref{eq:ariybc1B}-\ref{eq:ariybc4B}), with $\nu_B = \nu_A$, and edge extension $\alpha = \nu_A T/Y$ ! 
This observation immediately 
explains our numerical result: 
the planar stress field of the original problem (model A) is identical (up to a constant, purely uniaxial stress, $\sxx = T, \syy=\sxy=0$) 
to the stress field in a sheet whose short edges are pulled outward, and no longitudinal tension.
\subsubsection{The origin of transverse compression}
The above discussion reveals that the origin of transverse compression in a longitudinally-stretched sheet whose short edges are clamped is the ``edge-inducd'' potential $\Phi_e (x,y)$ in the decomposition~(\ref{eq:decompose-A}), or equivalently the Airy potential of our model B. It is thus natural 
to seek a solution using a basis of eigenfunctions of the bi-harmonic equation: 
\begin{subequations}
\begin{gather}
\Phi^{i,c}(x,y) = e^{-p_i\tfrac{x}{W}} \cos(p_i \frac{y}{W})  \ ,  \nonumber \\ 
\Phi^{i,s}(x,y) = e^{-p_i\tfrac{x}{W}} \frac{y}{W} \sin(p_i \frac{y}{W})   \ , 
\label{eq:basis-1}
\end{gather}
where $\{p_i \}_{i=1}^{\infty}$ is a discrete set of (generally complex) eigenvalues, which must be determined through the BCs~(\ref{eq:ariybc1B}-\ref{eq:ariybc4B}), and the symmetry: $\Phi(x,y) = \Phi(x,-y)$ has been exploited. 

Using the basis functions~(\ref{eq:basis-1}), we can express the Airy potential 
$\Phi_e (x,y)$ 
as: 
\begin{equation}
\Phi_e(x,y) = \text{Re}\! \left\{ \! \sum_i  C_i \!\cdot\! \left(\Phi^{i,c}(x,y)  + A_i \Phi^{i,s}(x,y) \right) \! \right\} , 
\label{eq:basis-2}
\end{equation}
where the sequences $\{p_i \} , \{A_i \}$, are determined by the free-edge BCs~(\ref{eq:ariybc1B},\ref{eq:ariybc2B}), yielding after some elementary manipulations: 
\begin{gather}
    p_i +\sin p_i = 0 \label{eq:Aipi1} \\
    A_i = -2 \cot \tfrac{p_i}{2} \label{eq:Aipi2} \ , 
\end{gather}
and the sequence $\{C_i \}$ is determined from the two extended-edge BCs~(\ref{eq:ariybc3B}, \ref{eq:ariybc4B}). Remarkably, Eq.~(\ref{eq:Aipi1}) reveals two facts on whose importance we will elaborate below: 
{\emph{(a)}} All eigenvalues $p_i$ are non-real numbers. 
{\emph{(b)}} the eigenvalues $p_i$ do not depend on the length-to-width ratio $\fLW$, nor on the Poisson ratio $\nu_B$. 

Evaluating the sequence $\{C_i \}$ requires an inverse Laplace transform \cite{Benthem1963}, which is a rather lengthy calculation, and furthermore, its implementation requires us to consider the limit of a semi-infinite sheet ($\fLW \to \infty$, whose extended short edge is at $x=0$). We will thus defer the details of this technical calculation to a separate publication, and report in Table I below only the values of the three eigenvalues $p_i$ with the smallest (positive) real parts, which govern the sum~(\ref{eq:basis-2}), together with the corresponding values of $A_i$ and $C_i$. 
\begin{center}
\begin{table}[h]
\begin{tabular}{| c | c | c |}\hline
         & $p_i$ & $A_i$ \\  \hhline{|===|}
   $i=1$ & $4.212+2.251j$ & $0.332 + 1.78j$ \\  
   $i=2$ & $10.713+3.103j$ & $0.168 + 1.94 j$ \\   
   $i=3$ & $17.073+3.551j$ & $0.111 + 1.97j$ \\  \hline
\end{tabular}

\begin{tabular}{| c | c | c |}\hline
     & $\nu_B = 0.32$ ($\times 10^{-3}$)  & $\nu_B = 0$ ($\times 10^{-3}$)\\ \hhline{|===|}
  $C_1$  & $109 - 213j$ & $131 - 229j$  \\  
  $C_2$  & $-22.9+ 13.5 j$ & $-26.1 + 11.7j$  \\  
 $C_3$  & $8.68 - 2.84j$ & $9.49- 1.52j$  \\  \hline
\end{tabular}
\caption{The values of the three leading poles ($p_i$), and corresponding pairs of coefficients ($A_i,C_i$)
in the expansion, Eq.~(\ref{eq:basis-2}), for the Airy function. Note that the pole $p_i$ and coefficients $A_i$ 
are independent on the Poisson ratio.}
\end{table}
\end{center}
Figures~\ref{fig:schematic}C and \ref{fig:schematic}D show that approximating the Airy potential through the first term in the sum (\ref{eq:basis-2}), {\emph{i.e.}} 
\begin{gather}
\Phi_e(x,y) \approx \nonumber \\
{\text{Re}}\{ C_1  e^{\!-p_1\tfrac{x}{W}} [ \cos (p_1 \frac{y}{W}) \!\! -\!\! 2\cot(\frac{p_1}{2}) \frac{y}{W} \sin(p_1 \frac{y}{W}) ]\} , \label{eq:approx-Phi-1}
\end{gather}
matches already very well the transverse stress obtained in the numerical solution at the vicinity of 
the short edges of a sheet with $\fLW = 8$. Furthermore, the analytic solution of the edge-induced Airy function $\Phi_e(x,y)$, 
given by Eq.~(\ref{eq:basis-2}) with the numerical values of $\{p_i,A_i,C_i \}$ in Table 1, 
provides some valuable insights into the mechanism by which transverse compression develops in an elongated sheet.  \\

$\bullet$ First, Table 1 indicates that the dependence on Poisson ratio, which stems only from the 
sequence $\{C_i\}$, is very weak. 
This observation, which has been noted already in our numerical analysis, substantiates the rationale 
illustrated in Fig.~\ref{fig:mapping} -- the primary cause of transverse compression is 
the extension of the short edges relative to the bulk, rather than  
the Poisson ratio of the sheet. \\ 

\newcommand{\im}{\text{Im\,}}
\newcommand{\re}{\text{Re\,}}
$\bullet$ Second, the unavoidable presence of transversely compressed zone in a sufficiently long sheet is a direct consequence of the fact that all eigenvalues $\{p_i\}$, 
namely, roots of Eq.~(\ref{eq:Aipi1}), are complex. The implication is revealed by evaluating $\syy$ from the approximated Airy potential~(\ref{eq:approx-Phi-1}) along the centerline ($y=0$) of a semi-infinite sheet: 
\begin{gather}
    \syy (x,y=0)  \propto e^{-p_1^{(r)} \cdot \tfrac{x}{W}} \cos[p_1^{(i)} \cdot \tfrac{x }{W} + g]  \nonumber \ ,  \\
    g = \tan^{-1}\left ( \frac{C_1^{(i)} {p_1^{(r)}}^2 + 2 C_1^{(r)} p_1^{(i)} p_1^{(r)} -C_1^{(i)} {p_1^{(i)}}^2}{C_1^{(r)} {p_1^{(i)}}^2 + 2 C_1^{(i)} p_1^{(i)} p_1^{(r)} - C_1^{(r)} {p_1^{(r)}}^2} \right )
    \label{eq:approx-syy-1}
\end{gather}
where the superscripts $^{(i)}$ and $^{(r)}$ refer to the imaginary and real parts, respectively. It is evident from the first line of Eq.~(\ref{eq:approx-syy-1}) that the imaginary component of the root, $p_1^{(i)} \neq 0$, gives rise to negative ({\emph{i.e.}} compressive) transverse stress at $-\tfrac{1}{2}L + d^*W <x < -\tfrac{1}{2}L+d_m^*W$ and $\tfrac{1}{2}L - d_m^*W <x < \tfrac{1}{2}L-d^*W$, where: 
\begin{gather} 
d^*  =(\pi/2 - g)/p_1^{(i)} \approx 0.646\nonumber \\
d_m^* = d^* + \tan^{-1}(p_1^{(i)}/p_1^{(r)})/p_1^{(i)} \approx 0.864 \ . 
\label{eq:approx-syy-2}
\end{gather} \\


$\bullet$ Third, 
Eq.~(\ref{eq:approx-syy-2}), indicates that 
the response of a 
rectangular sheet whose short edges are extended relative to the bulk, can be classified into three types, depending solely 
on the aspect ratio, $\fLW$:    \\

{\bf (I)} For $\fLW < 2 d^*$ there is no transverse compression. Here, the transverse stress, which is obviously tensile at the far edges ($x = \pm \tfrac{1}{2} L$), does not have enough room to vary significantly, hence the whole sheet is under pure (biaxial) tension. \\  

{\bf (II)} For $2 d^* <\fLW < 2 d_m^*$  there is a single transversely-compressed zone located around the center of the sheet. Here, the sheet is sufficiently elongated such that the transverse stress has enough room to approach negative values away from the tensed edges, but not to overturn and decay to zero. Hence, the two compressive zones, generated by each of the tensed edges, are merged into a single one.   \\ 

{\bf (II)} If $\fLW > 2 d_m^*$ the sheet is long enough such that there are two transversely-compressed zones, each of them starts at a distance $d^*W$ from a tensed edge, and extends over a length $\propto W$. \\




Figure 3 shows the transverse stress profile, obtained from our simulations for several representative values of the aspect ratio $L/W$, supporting the above classification into three regimes. We note that the actual values of $d^*$ and $d_m^*$ obtained from our simulations are rather close, but not identical, to the theoretical prediction, Eq.~(\ref{eq:approx-syy-2}). An obvious reason for this discrepancy is that
the values of $d^*,d_m^*$, reported in Eq.~(\ref{eq:approx-syy-2}), are obtained from an analytic solution of the transverse stress in a semi-infinite sheet ({\emph{i.e.}} $\tfrac{L}{W} \to \infty$), and we may thus expect corrections of $O(\tfrac{W}{L})$ to this predictions. From this viewpoint,  Fig.~3 indicates that those corrections to are in fact surprisingly small. Thus, while the above classification has been noted before by numerous workers ({\emph{e.g.}} \cite{Nayya11}), our analytic approach elucidates the 
origin of this classification through the complex values of the eigenvalues $\{p_i\}$ of the bi-harmonic equation under the BCs~(\ref{eq:ariybc1B},\ref{eq:ariybc1B}). \\

$\bullet$ Finally, we note that employing the basis functions (\ref{eq:basis-1}) yields a 
rapidly converging sequence and thereby the compact expression (\ref{eq:approx-Phi-1}) that describes 
quantitatively the stress field throughout the whole sheet. This global approach, which has been employed broadly for
solving the bi-harmonic equation in viscous fluid mechanics and linear elasticity problems  \cite{Benthem1963,Spence83} 
is thus advantageous to an approximation using ``corner functions'' \cite{Chopin18} that does not explain the emergence of transverse compression. 










 
\end{subequations}

\subsubsection{Analogy to Moffat eddies in a "driven cavity" flow}
\begin{figure}
\centering
\includegraphics[width=\linewidth]{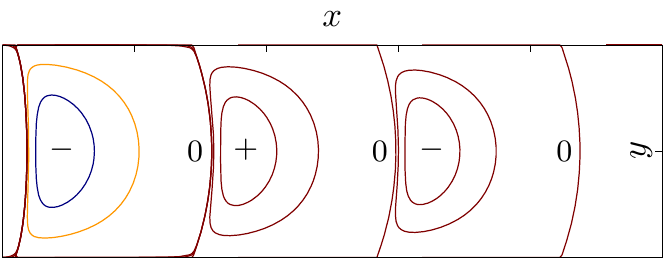}\\
\includegraphics[width=\linewidth]{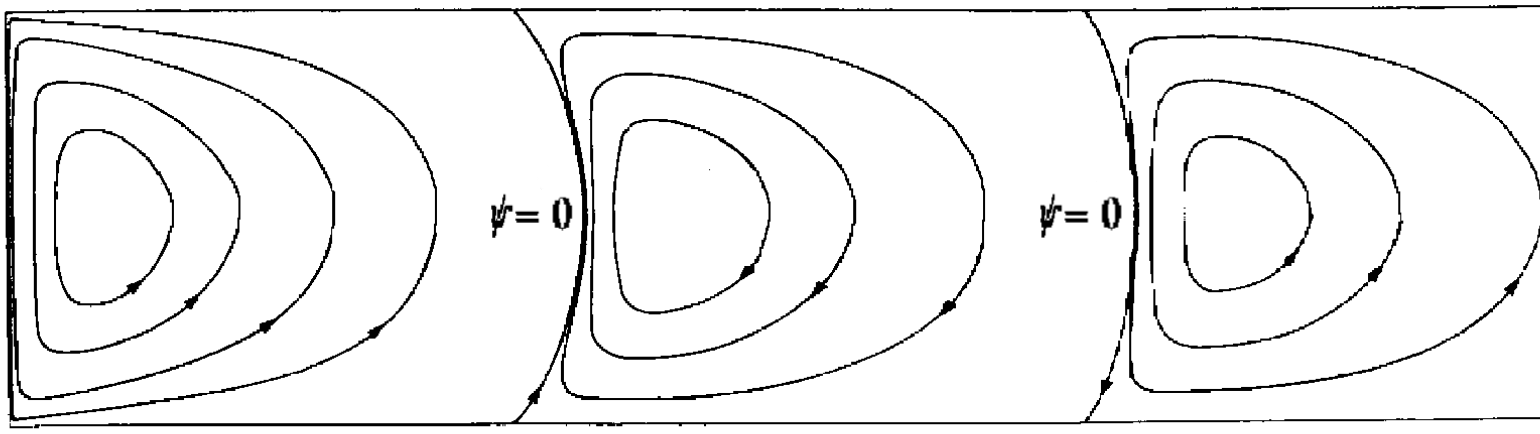}
\caption{Top: Contour plot of Airy potential $\Phi$ that describes the planar stress in a semi-infinite Hookean strip 
(\ref{eq:approx-Phi-1}). The symbols ``$+$" and ``$-$" 
indicate the sign of 
$\Phi$. 
Bottom: Moffatt eddies structure in a 2D Stokes flow in a driven rectangular cavity 
\cite{Shankar2000}. 
The viscous stress (and consequently the velocity field) in an incompressible Newtonian fluid is derived from a stream function that satisfies the bi-harmonic equation, similarly to the way by which the elastic stress in a solid Hookean sheet is derived from the Airy potential.} 
 \label{fig:Eddies}
\end{figure}

We have seen that the existence of complex eigenvalues of the bi-harmonic equation (\ref{eq:ariy}) for the edge-induced Airy potential $\Phi_e$ gives rise to a non-monotonic transverse stress $\syy = \partial_{xx} \Phi_e$, and consequently to transversely-compressed zones. It is useful to point out an analogy between this (arguably non-intuitive) effect and a classical phenomenon in fluid mechanics, known as ``Moffatt eddies'' \cite{Moffatt64}. Considering a class of two-dimensional (2D) viscous flows generated by the motion of rigid boundaries, Moffatt showed that solutions of the bi-harmonic equation, which describes the stream function of 2D Stokes flows, may be characterized by complex eigenvalues. A notable implication of this basic observation is the emergence of eddies in the ``driven-cavity'' set-up, whereby a rigid plate is moving at a constant velocity, dragging the surface of a viscous fluid enclosed in a deep cavity \cite{Shankar2000} (Fig.~\ref{fig:Eddies}). The formation of these eddies is intimately related to the complex eigenvalues that govern the stream function. The real (negative) part of the eigenvalues reflects the intuitive fact that the magnitude of the viscous stress (and consequently the speed) decays away from the driven surface, whereas the imaginary part implies that the decaying stress is nevertheless non-monotonic, and consequently an alternating direction of the velocity.

Thus, the transverse compression induced in a solid sheet by a relative extension of an edge with respect to the bulk may be viewed as an ``elastic analog'' of Moffatt eddies in a 2D driven-cavity viscous flow, providing a notable example of the Stokes-Rayleigh analogy between the mechanical equilibrium of Hookean solids and the viscous flow of Newtonian fluids.            

\section{Buckling instability} 
\label{sec:buckling}
When subjected to compressive loads,  
slender solid bodies become unstable,    
``trading'' a highly-energetic strain (averaged through the body's cross section), which is penalized by the stretching modulus $Y$, with curvature of the body's mid-plane, which is 
penalized by the bending modulus, $B \sim Y t^2$. The primary question we seek to address here is how the instability mode and the threshold value of the control parameter (longitudinal tension $T$ in model A and edge extension $\alpha $ for model B) depend on the sheet thickness $t$.   

Before studying this instability in our problem, where the planar stress is nonuiform 
(Figs.~\ref{fig:schematic}c,\ref{fig:schematic}d,\ref{fig:aspect-ratio}),   
it is useful to recall the basic example of a rectangular sheet under uniaxial compressive load (Fig.~\ref{fig:schematic-2}), where the planar stress is uniform and purely compressive ($\syy = -\sigma_0 <0 \ , \ \sxx=\sxy=0$). 

\subsection{Instability under uniform compression}
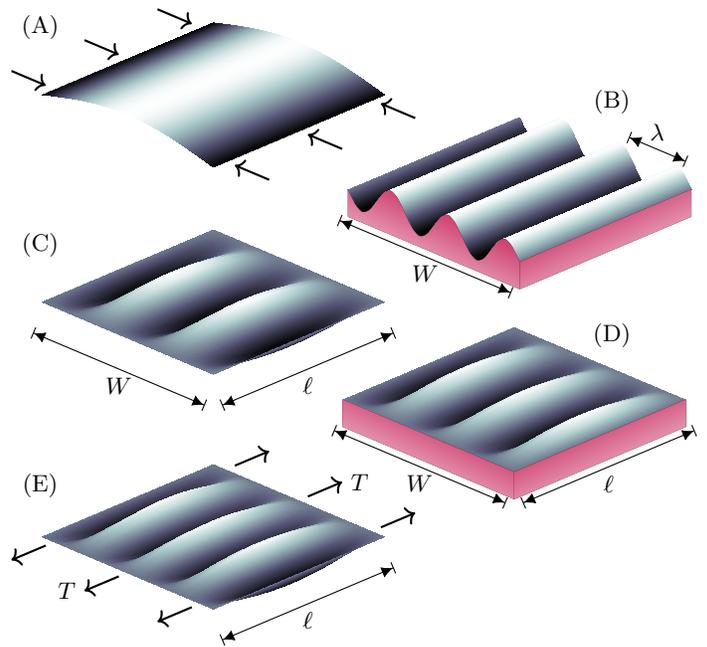
\begin{figure}
\begin{tikzpicture}
    \node at (-2., 1) {\begin{tikzpicture}
    \begin{axis}[
    hide axis,
    view={-45}{25},
    shader=interp,
    unit vector ratio*=1 1 1,
    ymin=-1,
    ymax=5,
    colormap={bw}{
        gray(0cm)=(0);
        gray(1cm)=(1);
    },
    colormap/bone,
    ]
        \addplot3[surf, domain=0:4, 
                  samples=2,
                  samples y=100
                 ] {0.3* sin(pi/4 * deg(y) )};
        \pgfplotsinvokeforeach{0.3, 2, 3.7} {
            \draw[->, thick] (#1, -1, 0) -- (#1, -0.2, 0); 
            \draw[->, thick] (#1, 5, 0 ) -- (#1, 4.2, 0);
        };
    \end{axis}
\end{tikzpicture}}; 
    \node at (2, -0.5) {\begin{tikzpicture}
        \begin{axis}[
        hide axis,
        view={-45}{25},
        unit vector ratio*=1 1 1,
        ymin=-1,
        ymax=5,
        zmin=-1.5,
        zmax=1,
        colormap/bone,
        xlabel=$x$,
        ylabel=$y$,
        ]
                \addplot3[surf,shader=interp, domain=0:4, 
                        samples=2,
                       samples y=100
                      ] {0.3* sin(2 * pi * 3 /4 * deg(y) )};
            \fill[top color=purple, bottom color=purple!10, 
            shading angle=-25] (0,0,-0.5) -- plot[variable=\y,domain=0:4, smooth] (0,\y,{0.3* sin(2 * pi * 3 /4 * deg(\y) )}) -- (0,4,-0.5) --
            cycle;
            \fill[top color=purple, bottom color=purple!10, 
            shading angle=25
            ] (0,0,-0.5) -- (0,0,0) -- (4, 0, 0) -- (4, 0, -0.5) -- cycle;
            \draw[{|Latex[width=1.5mm,length=1.5mm]}-{Latex[width=1.5mm,length=1.5mm]|}] (-0.2, 0, -0.5) -- (-0.2, 4, -0.5) node[midway, below] {$W$}; 
            \draw[{|Latex[width=1.5mm,length=1.5mm]}-{Latex[width=1.5mm,length=1.5mm]|}] (4.2, 0.33, 0.33) -- (4.2, 1.66, 0.33) node[midway, above] {$\lambda$};
        \end{axis}
\end{tikzpicture}};
    \node at (-2., -2) {\begin{tikzpicture}
    \begin{axis}[
        hide axis,
        view={-45}{25},
        unit vector ratio*=1 1 1,
        ymin=-1,
        ymax=5,
        zmin=-1.5,
        colormap/bone,
        xlabel=$x$,
        ylabel=$y$,
        ]
            \addplot3[surf,shader=interp, domain=0:4, 
                    ] {-0.3* sin(pi/4 * deg(x)) * 
                    sin(pi * 5 /4 * deg(y) )};
            \draw[{|Latex[width=1.5mm,length=1.5mm]}-{Latex[width=1.5mm,length=1.5mm]|}] (0,-0.2,-0.5) -- (4, -0.2, -0.5) node[midway, below] {$\ell$};
            \draw[{|Latex[width=1.5mm,length=1.5mm]}-{Latex[width=1.5mm,length=1.5mm]|}] (-0.2, 0, -0.5) -- (-0.2, 4, -0.5) node[midway, below] {$W$}; 
        \end{axis}
\end{tikzpicture}};
    \node at (2., -3.3) {\begin{tikzpicture}
    \begin{axis}[
        hide axis,
        view={-45}{25},
        unit vector ratio*=1 1 1,
        ymin=-1,
        ymax=5,
        zmin=-1.5,
        colormap/bone,
        xlabel=$x$,
        ylabel=$y$,
        ]
            \addplot3[surf,shader=interp, domain=0:4, 
                    ] {0.3* sin(pi/4 * deg(x)) * sin(2 * pi * 3 /4 * deg(y) )};
            \fill[top color=purple, bottom color=purple!10, 
            shading angle=-25] (0,0,-0.5) -- (0, 0, 0) -- (0,4,0) -- (0,4,-0.5) --
            cycle;
            \fill[top color=purple, bottom color=purple!10, 
            shading angle=25
            ] (0,0,-0.5) -- (0,0,0) -- (4, 0, 0) -- (4, 0, -0.5) -- cycle;
            \draw[{|Latex[width=1.5mm,length=1.5mm]}-{Latex[width=1.5mm,length=1.5mm]|}] (0,-0.2,-0.5) -- (4, -0.2, -0.5) node[midway, below] {$\ell$};
            \draw[{|Latex[width=1.5mm,length=1.5mm]}-{Latex[width=1.5mm,length=1.5mm]|}] (-0.2, 0, -0.5) -- (-0.2, 4, -0.5) node[midway, below] {$W$}; 
        \end{axis}
\end{tikzpicture}};
    \node at (-2., -5)
    {\begin{tikzpicture}
    \begin{axis}[
        hide axis,
        view={-45}{25},
        unit vector ratio*=1 1 1,
        ymin=-1,
        ymax=5,
        xmin=-1,
        xmax=5,
        zmin=-1.5,
        colormap/bone,
        xlabel=$x$,
        ylabel=$y$,
        ]
            \addplot3[surf,shader=interp, domain=0:4, 
                    ] {-0.3* sin(pi/4 * deg(x)) * 
                    sin(pi * 7 /4 * deg(y) )};
            \pgfplotsinvokeforeach{0.3, 2, 3.7} {
                \draw[<-, thick] (-1, #1, 0) -- (-0.2, #1, 0); 
                \draw[<-, thick] (5, #1, 0 ) -- (4.2, #1, 0);
            };
            \node[left] at (-1, 2, 0) {$T$};
            \node[right] at (5, 2, 0) {$T$};
            \draw[{|Latex[width=1.5mm,length=1.5mm]}-{Latex[width=1.5mm,length=1.5mm]|}] (0,-0.2,-0.5) -- (4, -0.2, -0.5) node[midway, below] {$\ell$};
        \end{axis}
\end{tikzpicture}};
    \node at (-4.3, 1.9) {(A)};
    \node at (-4.3, -1) {(C)};
    \node at (-4.3, -4.2) {(E)};
    \node at (3.3, 0.9) {(B)};
    \node at (3.3, -2.2) {(D)};
\end{tikzpicture}

\caption{A schematic of a rectangular sheet under uniaxial compressive load, $\syy = -\sigma_0$. 
(A) A classic version of the Euler buckling instability, where the edges, $x = \pm\ell/2$, are free, and the unstable mode consists of a single undulation ($\lambda \sim W$), regardless of the length $\ell$. (B) Attachment to an elastic substrate (``Winkler foundation'') of stiffness $\Ksub$ enhances resistance to undulations and thereby the threshold $\sigma_c$, and reducing the wavelength.      
(C-D) If the amplitude is suppressed at the edges $x = \pm\ell/2$, the instability mode is affected by the length $\ell$, such that the threshold $\sigma_c$ is increased in comparison to (a) and the near-threshold undulation wavelength $\lambda_c$ is decreased. 
For $\ell \ll W$ the near-threshold pattern consists of periodic undulations of wavelength $\lambda_c \sim \ell$. (E) 
The presence of longitudinal tension, $\sxx = T$, acts as an ``effective substrate'' of stiffness $K \sim T/\ell^2$ \cite{Cerda-Mahadevan2003}, affecting further reduction of the wavelength and correspondingly enhancement of the threshold value $\sigma_c$.}
\label{fig:schematic-2}
\end{figure}

The most elementary type of an elastic instability under a uniform uniaxial compressive load, $\syy = -\sigma_0$, 
is exhibited by an unsupported rectangular sheet, of thickness $t$ and width $W$, whose edges, $x=\pm \tfrac{1}{2}\ell$, are free (Fig.~\ref{fig:schematic-2}a). In a popular explanation of this instability, known as ``Euler buckling'', the deflected state of the sheet is modeled as an Euler {\emph{elastica}} -- a strainless deformation that fully converts compression into out-of-plane deflection -- and its bending energy ($\propto B\tfrac{\sigma_0}{Y} \tfrac{1}{W^2}$) is compared with the strain energy of the planar state ($\propto \tfrac{\sigma_0^2}{Y}$). 
We note, however, that this common explanation is a 
``far from threshold'' reasoning, which does not capture the physics at the vicinity of a continuous (supercritical) bifurcation.  
Instead, we provide here another explanation for the Euler instability, addressing it as one that emerges from a standard supercritical 
bifurcation, in accord with the near-threshold analysis that is the focus of the current paper.


Consider then 
some given value of $\sigma_0$, and assume an undulatory deviation
$\zeta(x,y) = A \cdot g_\lambda(y)$ from the planar state,  where the amplitude $A$ is \emph{infinitesimal}, and $g_\lambda(y)$ is some function that undulates over a scale $\lambda$. Generally, $g_\lambda(y)$ is an eigenfunction of the elastic energy functional, linearized around the planar state, and for this highly symmetric problem it is simply sinusoidal. Nevertheless, we prefer to keep a more general terminology in order to highlight the commonality with our original model problem, 
which is far less symmetric.   
Such a perturbation reduces slightly 
the strain in the planar state: 
$\varepsilon_{yy} \to -\tfrac{\sigma_0}{Y} + C_1 \cdot (\tfrac{A}{\lambda})^2$, where $C_1>0$ is some numerical constant.
For an infinitesimal $A$, the strain energy is reduced (from the planar-state) by a value  
$ \propto \ \sigma_0 (\tfrac{A}{\lambda})^2$, whereas the bending energy (which is obviously zero at the planar state) is increased by a value $\propto \ B(\tfrac{A}{\lambda^2})^2$, and one readily finds that such a perturbation of the planar state becomes favorable once the compressive load exceeds a threshold, $\sigma_0^* (\lambda)\sim B/\lambda^2$. Since the wavelength $\lambda$ is limited by the width $W$, we find that the planar state first becomes unstable to undulations at the largest possible wavelength, 
when the exerted compressive load exceeds a threshold value, $\sigma_c = \sigma_0^*(W)$: 
\begin{equation}
\text{Euler \ buckling:}  \ \ 
\sigma_c 
\sim \frac{B}{W^2} \sim Y (\frac{t}{W})^2 \  ;  \  \lambda \sim W \ .  
\label{eq:Euler-1}
\end{equation}

Let us consider now two other variants 
of the instability under uniform, uniaxial compressive load. The first variant, known as the Winkler model and depicted in Fig.~\ref{fig:schematic-2}b, consists of a sheet attached to an elastic substrate, which penalizes vertical displacement by an energy (per area), $\tfrac{1}{2} \Ksub \zeta^2$. The additional energy cost for undulations implies that, for a given $\lambda$, 
a planar state becomes unstable only if $\sigma_0$ 
exceeds an enhanced threshold value, $\sigma_0^*(\lambda) \sim B/\lambda^2 + \Ksub \lambda^2 $. Consequently, 
we find that the planar state first becomes unstable to undulations at a wavelength, $\lambda_c \sim (B/\Ksub)^{1/4}$, which may be $\ll W$, 
when the exerted compressive load exceeds a threshold value, $\sigma_c \sim \sqrt{B \Ksub} \sim \sqrt{Y\Ksub} \cdot t$. 
These two features -- a wavelength that exhibits strong dependence on the sheet thickness ({\emph{i.e.}} $\lambda \sim t^{\beta_\lambda}$ with $\beta_\lambda >0$) and an enhanced threshold for destabilizing a planar state ({\emph{i.e.}} $\sigma_c/Y \sim t^{\beta_\sigma}$ with $\beta_\sigma <2$) are hallmarks of wrinkling phenomena, demarcating them from the standard version of Euler buckling instability.       

A second variant of instability under uniform, uniaxial compressive load, is depicted in Figs.~\ref{fig:schematic-2}c-d. 
Here, the amplitude is suppressed at the edges $x=\pm \tfrac{1}{2}\ell$, such that 
a shape $A \cdot \zeta_\lambda(x,y)$ that undulates over a characteristic scale $\lambda$ along the compressive axis ($\hat{y}$), must vary also along the $\hat{x}$ axis, thereby being penalized also by the bending cost of  
the corresponding curvature, $\propto A/\ell^2$. 
For a given $\lambda$, 
a planar state becomes unstable only if the copressive load exceeds, $\sigma_0^* (\lambda) \sim B(1/\lambda^2 + \lambda^2/\ell^4)$, and if $\ell \ll  W$ we find that the planar state first becomes unstable to undulations of wavelength, $\lambda_c \sim \ell$, at a threshold, $\sigma_c \sim B/\ell^2 \sim Y (\tfrac{t}{\ell})^2$. Furthermore, if the sheet is subjected also to a tensile load $\sxx = T$ along the longitudinal ($\hat{x}$) axis (Fig.~\ref{fig:schematic-2}e), there will be yet another energetic penalty for undulations, $\sim T (\tfrac{A}{\ell})^2$, which is analogous to the energy implied by an actual (Winkler) substrate, $\Ksub \propto T/\ell^2$. 

Putting together the effects of a real substrate, amplitude-suppressing boundaries, and longitudinal tension,   
we find that, for a given wavelength $\lambda$, the planar state of a rectangular sheet under uniform compression in the transverse ($\hat{y}$) axis becomes unstable if the compressive load $\sigma_0$ exceeds:     
\begin{subequations}
\begin{gather}
\sigma_0^*(\lambda) \sim B\frac{1}{\lambda^2} + (\Ksub +  T\frac{1}{\ell^2} + B\frac{1}{\ell^4}) \lambda^2   \ , 
\label{eq:gen-1}
\end{gather}
and hence the instability is characterized by a wavelength  
\begin{gather} 
\lambda_c \sim \min \left(W \ ,  \ (\frac{B}{\Ksub +  T \ell^{-2}  + B \ell^{-4} })^{1/4} \right) 
\label{eq:gen-2}
\end{gather}
and occurs as the compressive load exceeds a threshold 
\begin{gather}
\sigma_c \sim  \max \left(BW^{-2}  ,  \sqrt{B (\Ksub +  T \ell^{-2}  + B \ell^{-4})}      
\right)  . 
\label{eq:gen-3}
\end{gather}
\end{subequations}



\subsection{Why is the instability buckling-like ?}
Let us turn back now to our problem -- where the transverse stress induced by the relative edge in nonuniform  (Figs.~\ref{fig:schematic}c,\ref{fig:schematic}d,\ref{fig:aspect-ratio}), 
namely $\syy (x,y)$ varies along both $\hat{x}$ and $\hat{y}$ axes. One can still perform a linear stability analysis of the planar stress to 
infinitesimal deflections,  $\zeta(x,y) =  A\cdot g_\lambda(x,y)$, which undulate with a characteristic wavelength $\lambda$ along the $\hat{y}$ axis and an infinitesimal amplitude $A$. However, the lack of translation symmetry of the planar state implies that the eigenfunctions, $g_\lambda(x,y)$, of the corresponding (linearlized) energy functional are not simply sinusoidal Fourier modes. Nevertheless, as we explain below the physical mechanisms that determine the critical wavelength $\lambda_c$ and the threshold $\sigma_c$ for the uniform compression problem, Eqs.~(\ref{eq:gen-1}-\ref{eq:gen-3}), are analogous to those that govern the instability of the nonuniform planar stress in our problem, allowing us 
to gain valuable insights. 

Let us consider first model B, where the magnitude $\sigma_0 \propto \alpha \cdot Y$ of the transverse compressive stress is induced directly by the edge extension parameter, $\alpha$, and the stretching modulus $Y$. Here, there is no longitudinal tension ($T=0$), and -- since our sheet is unsupported ({\emph{i.e.}} $\Ksub = 0$) -- nor there is a real substrate effect. 
Since the transverse compression in the planar state is limited to a narrow zone in the sheet, the 
compressive stress $\sigma_0^*(\lambda)$ above which a undulation of wavelength $\lambda$ becomes favorable is 
subjected to the effect of 
an amplitude-suppressing boundaries discussed above (last term in Eq.~\ref{eq:gen-1}). However, since the length of the compressive zone is proportional to the sheet's width ({\emph{i.e.}} $\ell \propto W$), the overall effect on the critical wavelength $\lambda_c$ and threshold value $\sigma_c* \propto \alpha_c Y$ is inconspicuous, and we find the scaling: 
\begin{equation}
\text{model \ B:} \ \ \alpha_c \sim \frac{B}{YW^2} \sim (\frac{t}{W})^2 \ \ ; \ \ \lambda_c \sim W   \ .     
\label{eq:thereshold-11}
\end{equation}

Turning now to our original problem (model A), where the magnitude of the transverse compressive stress $\sigma_0 \propto T$,  
we recognize an additional contribution to $\sigma_0^*(\lambda)$, Eq.~(\ref{eq:gen-1}), {due to the energetic cost for deflection over a length $\ell \sim W$ along the tensile  axis. However, since $\sigma_0$ is also proportional to the longitudinal tension $T$, we find that the minimal value of $T$ for which Eq.~(\ref{eq:gen-1}) is satisfied is again realized when the wavelength $\lambda$ is a finite, thickness-independent } fraction of the sheet width, implying:
\begin{equation}
\text{model \ A:} \ \ T_c \sim \frac{B}{W^2} \sim Y\cdot (\frac{t}{W})^2 \ \ ; \ \ \lambda_c \sim W   \ .  
\label{eq:thereshold-12}
\end{equation}

Thus, notwithstanding the narrowness of the compressive zone and the presence of longitudinal tension in it, inspection of Eq.~(\ref{eq:Euler-1}) and Eqs.~(\ref{eq:thereshold-11},\ref{eq:thereshold-12}) reveals that the instability of the planar shape caused by relative edge extension
exhibits the typical behavior of the classic Euler buckling instability, namely, a thickness-independent critical ``wavelength'' $\lambda_c$ set by the sheet geometry, and a threshold load value that scales as the square power of the thickness-width ratio. 

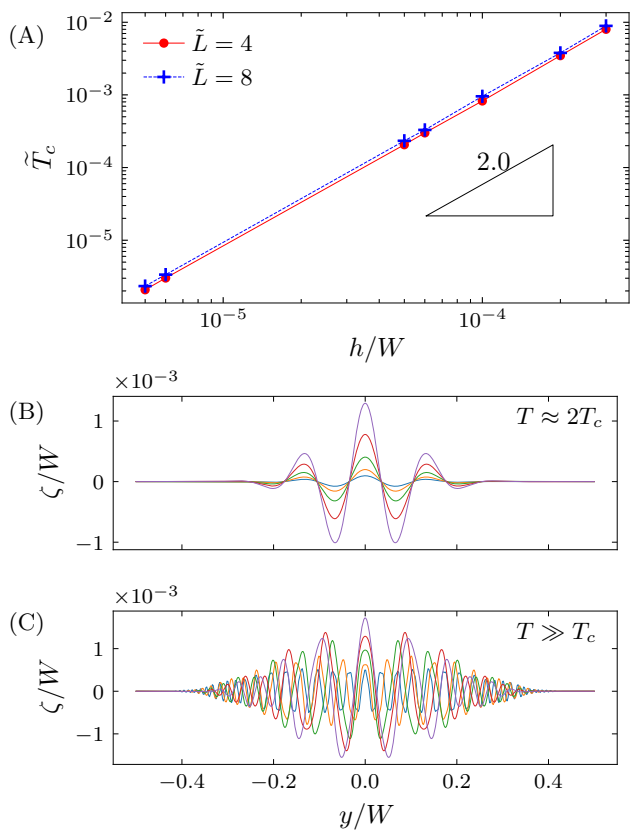
\begin{figure}
\centering
\begin{tikzpicture}
    \node at (0,0) {\includegraphics[]{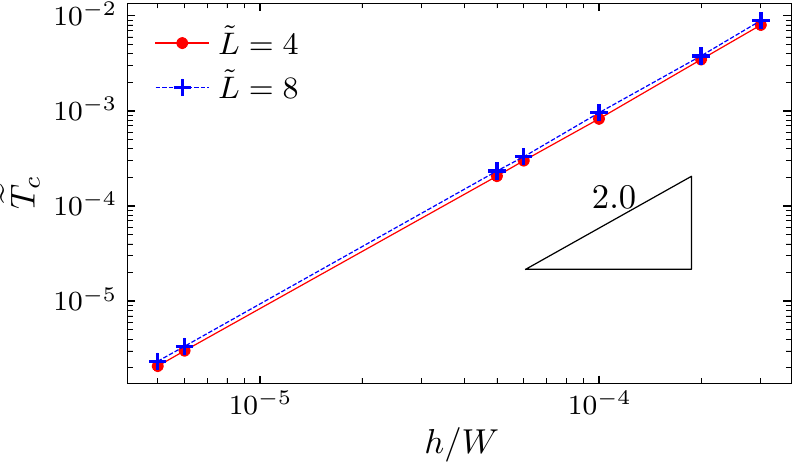}};
    \node at (0,-5.5) {\includegraphics[]{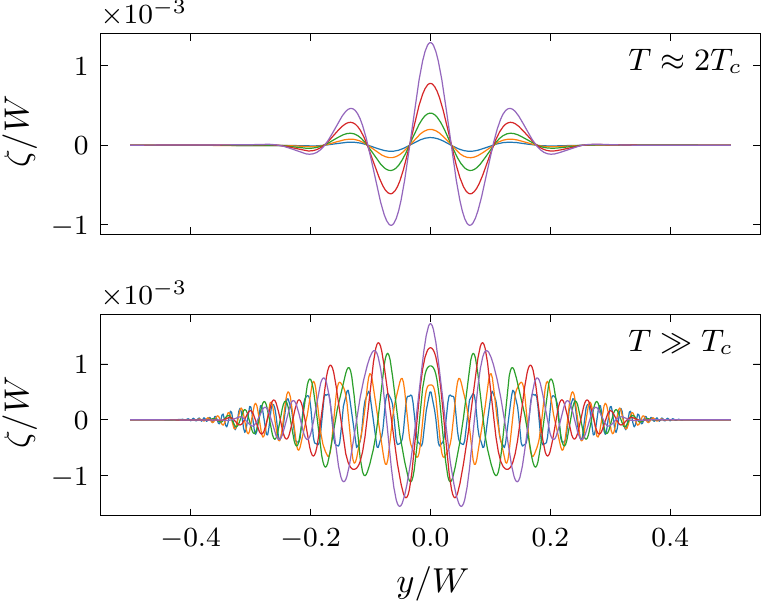}};
    \node at (-4, 2) {(A)};
    \node at (-4, -3) {(B)};
    \node at (-4, -5.8) {(C)};
\end{tikzpicture}

\caption{(A) The instability threshold $\tT_c$, of a long stretched rectangular sheet with clamped edges (model A with $L=8W$), 
as obtained from our SE simulations, plotted {\emph{vs.}} the thickness, $\tfrac{t}{W}$.
The threshold value is shown to be proportional to $(\tfrac{t}{W})^2$, with a proportionality constant that 
depends weakly on the aspect ratio $\tfrac{W}{L}$ for $L\gg W$. This result indicates that the instability is an Euler-like buckling, due to a compressed zone of width $\sim W$, where the compression is, $\syy \sim -T$ (see text). 
(B) The transverse profile of the shape (at $x\approx x_{max}$, where the compression is maximal), plotted close to threshold, indicates 
that the critical wavelength,
$\lambda_c$, is a finite, thickness-independent fraction of the sheet width,  
in accord with Euler buckling instability. 
(C) As the exerted tension $T$ is increased beyond $T_c$, the energetically-favorable wavelength $\lambda$ becomes smaller 
(in comparison to $\lambda_c$, and develop explicit dependence on the sheet thickness. 
This buckling-to-wrinkling trend is consistent with a transition from near-threshold to far-from-threshold behavior envisioned in \cite{Cerda-Mahadevan2003}. From top to bottom (looking at center $y/W = 0$), $T/T_c \approx 66,\ 118,\ 266,\ 1074,\ 4324$.} 
 \label{fig:threshold}
\end{figure}

Figure~\ref{fig:threshold} shows that the predicted buckling-like behavior, characterized by the scaling rules~(\ref{eq:thereshold-11},\ref{eq:thereshold-12}), is confirmed by our simulations. 
In Fig.~\ref{fig:threshold}a, threshold values ($T_c$ for model A and $\alpha_c$ for model B) were obtained for a range of sheet thicknesses 
by carefully probing intervals of the control parameters ($T$ and $\alpha$, respectively), and then plotted {\emph{vs.}} the aspect ratio $\tfrac{t}{W}$, showing an excellent agreement with the predicted scaling behavior. Apart from their identical scaling behavior, the threshold value of the dimensionless control parameter $T_c/Y$ in model A is larger than its counterpart $\alpha_c$ in model B, in accord with the enhanced resistance of the former to buckling, due to the effect of longitudinal tension in the transversely compressed zone. The enhanced resistance to undulations is reflected also in the near-threshold pattern (Fig.~\ref{fig:schematic}e-f). 
While both models exhibit near threshold a buckling ({\emph{i.e}} thickness-independent undulation) pattern, such that the wavelength $\lambda_c$ is a finite fraction of the width $W$, this fraction is smaller in model A (by a factor of $\approx \tfrac{1}{3}$) in comparison to its counterpart in model B.                          

\section{Beyond threshold}
\label{sec:transition}
Upon increasing the control parameter $T$ in model A 
substantially above its threshold value~(\ref{eq:thereshold-12}), 
our simulations (Fig.~\ref{fig:threshold}c) show that 
the near-threshold buckling pattern undergoes two dramatic changes. First, undulations expand (along the $\hat{x}$ axis) beyond the transversely-compressed zone of the planar state. Second, the characteristic wavelength $\lambda$ becomes substantially smaller than its threshold value $\lambda_c \sim W$. For model B, the analogous process of increasing $\alpha$ beyond threshold~(\ref{eq:thereshold-11}) does lead to expansion of the deflected zone, but not to any significant reduction in the characteristic undulation wavelength. 

A systematic study of this dramatic evolution from a buckling pattern to wrinkles in the (Hookean) far-from-threshold regime ($T_c \ll T \ll Y$) in model A, requires a detailed analysis of the strong effect of wrinkle formation on the stress field in the sheet, and its consequent departure from the planar stress. This non-perturbative effect, which 
revokes the perturbative approach underlying the near-threshold analysis in Sec.~\ref{sec:buckling}, will be the focal point of our subsequent paper. 
Here, we take a more    
%
%
%
heuristic approach to 
rationalize the qualitative distinction between models A and B, by generalizing the analysis in the preceding section beyond the near-threshold regime. 

Inspecting the considerations underlying the critical wavelength $\lambda_c $ (\ref{eq:gen-2}), 
one may notice that the only way in which the planar state is explicitly affecting the wavelength is through the length $\ell \sim W$ of the transversely-compressed zone. 
Assuming that even when the control parameter exceeds considerably the threshold value 
the wrinkle wavelength $\lambda$ is affected by the stress distribution 
through the length $\ell$ of the actual compressive zone, the rule (\ref{eq:gen-2}) can be generalized to: 
\begin{gather} 
\lambda \sim \min \left(W \ ,  \ (\frac{B}{\Ksub +  T \ell_*^{-2}  + B \ell_*^{-4} })^{1/4} \right)  \ , 
\label{eq:gen-20}
\end{gather}
where $\ell_*$ is the {\emph{actual}} length of the compressive zone (at a given, post-threshold value of the control parameter) rather its length in the planar state. 

With the generalized version~(\ref{eq:gen-20}) of the wavelength rule, one may immediately notice the difference between models A and B. In the former, the presence of longitudinal tension eventually dominates the wavelength, hence: 
\begin{equation}
\text{model \ A} \ (T \gg T_c) :  \ \  \lambda \sim (\frac{B}{T\ell_*^{-2}})^{1/4}  
\label{eq:Cerda-ext}
\end{equation} 
such that at a fixed value of $T/Y$, the wavelength $\lambda$ vanishes with the sheet thickness, signifying a transition from buckling ($\lambda \sim W$ at $T\approx T_c$) to wrinkling ($\lambda \sim t^{1/2} \ll W$ for $T \gg T_c$). The scaling rule of Cerda \& Mahadevan \cite{Cerda-Mahadevan2003} is obtained by assuming that the transversely-compressed zone extends throughout the whole sheet, {\emph{i.e.}} $\ell_* \sim L$ in Eq.~(\ref{eq:Cerda-ext}). In contrast, for model B, the absence of longitudinal  tension implies that the pattern does not undergo a similar buckling-to-wrinkling transition as the control parameter $\alpha$ exceeds the threshold value.    

The above heuristic argument deserves 
a healthy dose of skepticism. Why does the transversely-compressed zone expand when the sheet is driven away from threshold ? Why is it justified to approximate the energetic cost (per area) imposed on undulations by the longitudinal tension as $T (A/\ell_*)^2 $ ? To properly address these questions one has to consider the tension-field solution of this problem, which forms the basis for far-from-threshold analysis, and will be discussed in our subsequent paper.




\section{Summary}   
\label{sec:discussion} 

Focusing on the planar stress of a Hookean, rectangular-shaped sheet under unaxial, longitudinal tensile load, $\sxx \approx T$, we showed that the emergence of transversely-compressed zones stems from the extension of the pulled clamped edges 
relative to the bulk. Specifically, we showed that an identical profile of the transverse stress is realized by directly pulling the corners transversely without 
longitudinal tension (Fig.~\ref{fig:schematic}c). 
This observation evinces that the classic Poisson effect, namely, ``tension-induced {\emph{contraction}}'' of a solid sheet, must not be confused with ``tension-induced {\emph{compression}}" which underlies tensional wrinkling phenomena. The former is a {\emph{bulk effect}}, whereby transverse {\emph{strain}} ($\varepsilon_{yy} \sim -\nu T/Y$) emerges away from the edges in order to avoid compression; The latter is a {\emph{boundary effect}}, which can be eliminated by tailoring the boundary conditions ({\emph{e.g.}} unclamping the pulled edges), and hence should be referred to as ``edge-induced (transverse) compression''.   

Furthermore, we showed that edge-induced transverse compression stems from a non-monotonic decay of the Airy potential away from the edge, reflecting the effect of complex eigenvalues of the bi-harmonic equation, in analogy to Moffatt eddies in viscous driven cavity flow. 

Finally, we showed that localized, transversely compressed zones in the planar stress give rise to {\emph{buckling}} instability, with a critical wavelength $\lambda_c$ proportional to the sheet width $W$ and independent on the its thickness $t$, and a threshold tension $T_c \sim (\tfrac{t}{W})^2 Y$. Both of these relations mirror the classical Euler buckling, revealing the anticipation of Cerda \& Mahadevan \cite{Cerda-Mahadevan2003} that the commonly observed wrinkling pattern in this set-up, with a wavelength $\lambda$ that vanishes with $t$
(Fig.~\ref{fig:schematic}g,\ref{fig:threshold}c), cannot be described by a standard post-buckling theory that assumes moderate perturbation of the planar stress. A description of this wrinkling pattern through a far-from-threshold framework is the subject of a subsequent publication.  \\



We thank F. Brau, E. Cerda, J. Chopin, P. Damman, A. Kudroli, and N. Menon for valuable discussions. 
This research was funded by the National Science Foundation under grant DMR 1822439. Simulations were performed in the computing cluster of Massachusetts Green High Performance Computing Center (MGHPCC).









  


\bibliography{MS-I}

\begin{thebibliography}{30}%
\makeatletter
\providecommand \@ifxundefined [1]{%
 \@ifx{#1\undefined}
}%
\providecommand \@ifnum [1]{%
 \ifnum #1\expandafter \@firstoftwo
 \else \expandafter \@secondoftwo
 \fi
}%
\providecommand \@ifx [1]{%
 \ifx #1\expandafter \@firstoftwo
 \else \expandafter \@secondoftwo
 \fi
}%
\providecommand \natexlab [1]{#1}%
\providecommand \enquote  [1]{``#1''}%
\providecommand \bibnamefont  [1]{#1}%
\providecommand \bibfnamefont [1]{#1}%
\providecommand \citenamefont [1]{#1}%
\providecommand \href@noop [0]{\@secondoftwo}%
\providecommand \href [0]{\begingroup \@sanitize@url \@href}%
\providecommand \@href[1]{\@@startlink{#1}\@@href}%
\providecommand \@@href[1]{\endgroup#1\@@endlink}%
\providecommand \@sanitize@url [0]{\catcode `\\12\catcode `\$12\catcode
  `\&12\catcode `\#12\catcode `\^12\catcode `\_12\catcode `\%12\relax}%
\providecommand \@@startlink[1]{}%
\providecommand \@@endlink[0]{}%
\providecommand \url  [0]{\begingroup\@sanitize@url \@url }%
\providecommand \@url [1]{\endgroup\@href {#1}{\urlprefix }}%
\providecommand \urlprefix  [0]{URL }%
\providecommand \Eprint [0]{\href }%
\providecommand \doibase [0]{http://dx.doi.org/}%
\providecommand \selectlanguage [0]{\@gobble}%
\providecommand \bibinfo  [0]{\@secondoftwo}%
\providecommand \bibfield  [0]{\@secondoftwo}%
\providecommand \translation [1]{[#1]}%
\providecommand \BibitemOpen [0]{}%
\providecommand \bibitemStop [0]{}%
\providecommand \bibitemNoStop [0]{.\EOS\space}%
\providecommand \EOS [0]{\spacefactor3000\relax}%
\providecommand \BibitemShut  [1]{\csname bibitem#1\endcsname}%
\let\auto@bib@innerbib\@empty
\bibitem [{\citenamefont {Friedl}\ \emph {et~al.}(2000)\citenamefont {Friedl},
  \citenamefont {Rammerstorfer},\ and\ \citenamefont {Fischer}}]{Friedl2000}%
  \BibitemOpen
  \bibfield  {author} {\bibinfo {author} {\bibfnamefont {N.}~\bibnamefont
  {Friedl}}, \bibinfo {author} {\bibfnamefont {F.}~\bibnamefont
  {Rammerstorfer}}, \ and\ \bibinfo {author} {\bibfnamefont {F.}~\bibnamefont
  {Fischer}},\ }\href {\doibase https://doi.org/10.1016/S0045-7949(00)00072-9}
  {\bibfield  {journal} {\bibinfo  {journal} {Computers \& Structures}\
  }\textbf {\bibinfo {volume} {78}},\ \bibinfo {pages} {185 } (\bibinfo {year}
  {2000})}\BibitemShut {NoStop}%
\bibitem [{\citenamefont {Cerda}\ \emph {et~al.}(2002)\citenamefont {Cerda},
  \citenamefont {Ravi-Chandar},\ and\ \citenamefont
  {Mahadevan}}]{Cerda-Mahadevan2002}%
  \BibitemOpen
  \bibfield  {author} {\bibinfo {author} {\bibfnamefont {E.}~\bibnamefont
  {Cerda}}, \bibinfo {author} {\bibfnamefont {K.}~\bibnamefont {Ravi-Chandar}},
  \ and\ \bibinfo {author} {\bibfnamefont {L.}~\bibnamefont {Mahadevan}},\
  }\href {\doibase 10.1038/419579b} {\bibfield  {journal} {\bibinfo  {journal}
  {Nature}\ }\textbf {\bibinfo {volume} {419}},\ \bibinfo {pages} {579}
  (\bibinfo {year} {2002})}\BibitemShut {NoStop}%
\bibitem [{\citenamefont {Nayyar}\ \emph
  {et~al.}(2011{\natexlab{a}})\citenamefont {Nayyar}, \citenamefont
  {Ravi-Chandar},\ and\ \citenamefont {Huang}}]{Nayya2011}%
  \BibitemOpen
  \bibfield  {author} {\bibinfo {author} {\bibfnamefont {V.}~\bibnamefont
  {Nayyar}}, \bibinfo {author} {\bibfnamefont {K.}~\bibnamefont
  {Ravi-Chandar}}, \ and\ \bibinfo {author} {\bibfnamefont {R.}~\bibnamefont
  {Huang}},\ }\href {\doibase https://doi.org/10.1016/j.ijsolstr.2011.09.004}
  {\bibfield  {journal} {\bibinfo  {journal} {International Journal of Solids
  and Structures}\ }\textbf {\bibinfo {volume} {48}},\ \bibinfo {pages} {3471 }
  (\bibinfo {year} {2011}{\natexlab{a}})}\BibitemShut {NoStop}%
\bibitem [{Note1()}]{Note1}%
  \BibitemOpen
  \bibinfo {note} {``buckling'' is thus understood as a particular instance of
  ``wrinkling``, where the power $\alpha =0$}\BibitemShut {NoStop}%
\bibitem [{\citenamefont {Milner}\ \emph {et~al.}(1989)\citenamefont {Milner},
  \citenamefont {Joanny},\ and\ \citenamefont {Pincus}}]{Milner89}%
  \BibitemOpen
  \bibfield  {author} {\bibinfo {author} {\bibfnamefont {S.~T.}\ \bibnamefont
  {Milner}}, \bibinfo {author} {\bibfnamefont {J.~F.}\ \bibnamefont {Joanny}},
  \ and\ \bibinfo {author} {\bibfnamefont {P.}~\bibnamefont {Pincus}},\
  }\href@noop {} {\bibfield  {journal} {\bibinfo  {journal} {Euro. Phys.
  Lett.}\ }\textbf {\bibinfo {volume} {9}},\ \bibinfo {pages} {495} (\bibinfo
  {year} {1989})}\BibitemShut {NoStop}%
\bibitem [{\citenamefont {Bowden}\ \emph {et~al.}(1998)\citenamefont {Bowden},
  \citenamefont {Brittain}, \citenamefont {Evans}, \citenamefont {Hutchinson},\
  and\ \citenamefont {Whiteside}}]{Bowden98}%
  \BibitemOpen
  \bibfield  {author} {\bibinfo {author} {\bibfnamefont {N.}~\bibnamefont
  {Bowden}}, \bibinfo {author} {\bibfnamefont {S.}~\bibnamefont {Brittain}},
  \bibinfo {author} {\bibfnamefont {A.~G.}\ \bibnamefont {Evans}}, \bibinfo
  {author} {\bibfnamefont {J.~W.}\ \bibnamefont {Hutchinson}}, \ and\ \bibinfo
  {author} {\bibfnamefont {G.~M.}\ \bibnamefont {Whiteside}},\ }\href@noop {}
  {\bibfield  {journal} {\bibinfo  {journal} {Nature}\ }\textbf {\bibinfo
  {volume} {393}},\ \bibinfo {pages} {146} (\bibinfo {year}
  {1998})}\BibitemShut {NoStop}%
\bibitem [{\citenamefont {Cerda}\ and\ \citenamefont
  {Mahadevan}(2003{\natexlab{a}})}]{Cerda03}%
  \BibitemOpen
  \bibfield  {author} {\bibinfo {author} {\bibfnamefont {E.}~\bibnamefont
  {Cerda}}\ and\ \bibinfo {author} {\bibfnamefont {L.}~\bibnamefont
  {Mahadevan}},\ }\href@noop {} {\bibfield  {journal} {\bibinfo  {journal}
  {Phys. Rev. Lett.}\ }\textbf {\bibinfo {volume} {90}},\ \bibinfo {pages}
  {074302} (\bibinfo {year} {2003}{\natexlab{a}})}\BibitemShut {NoStop}%
\bibitem [{\citenamefont {Wagner}(1929)}]{Wagner29}%
  \BibitemOpen
  \bibfield  {author} {\bibinfo {author} {\bibfnamefont {H.}~\bibnamefont
  {Wagner}},\ }\href@noop {} {\bibfield  {journal} {\bibinfo  {journal} {Z
  Flugtechn Motorluftschiffahrt}\ }\textbf {\bibinfo {volume} {20}},\ \bibinfo
  {pages} {8} (\bibinfo {year} {1929})}\BibitemShut {NoStop}%
\bibitem [{\citenamefont {Stein}\ and\ \citenamefont
  {Hedgepeth}(1961)}]{stein61}%
  \BibitemOpen
  \bibfield  {author} {\bibinfo {author} {\bibfnamefont {M.}~\bibnamefont
  {Stein}}\ and\ \bibinfo {author} {\bibfnamefont {J.~M.}\ \bibnamefont
  {Hedgepeth}},\ }\href@noop {} {\emph {\bibinfo {title} {Analysis of Partly
  Wrinkled Membranes}}},\ \bibinfo {type} {Tech. Rep.}\ (\bibinfo
  {institution} {NASA},\ \bibinfo {year} {1961})\BibitemShut {NoStop}%
\bibitem [{\citenamefont {Mansfield}(1989)}]{MansfieldBook}%
  \BibitemOpen
  \bibfield  {author} {\bibinfo {author} {\bibfnamefont {E.~H.}\ \bibnamefont
  {Mansfield}},\ }\href@noop {} {\emph {\bibinfo {title} {The Bending and
  Stretching of Plates}}}\ (\bibinfo  {publisher} {Cambridge University
  Press},\ \bibinfo {year} {1989})\BibitemShut {NoStop}%
\bibitem [{\citenamefont {Pipkin}(1986)}]{Pipkin86}%
  \BibitemOpen
  \bibfield  {author} {\bibinfo {author} {\bibfnamefont {A.~C.}\ \bibnamefont
  {Pipkin}},\ }\href@noop {} {\bibfield  {journal} {\bibinfo  {journal} {IMA J.
  Appl. Math.}\ }\textbf {\bibinfo {volume} {36}},\ \bibinfo {pages} {85}
  (\bibinfo {year} {1986})}\BibitemShut {NoStop}%
\bibitem [{\citenamefont {Steigmann}(1990)}]{Steigman90}%
  \BibitemOpen
  \bibfield  {author} {\bibinfo {author} {\bibfnamefont {D.~J.}\ \bibnamefont
  {Steigmann}},\ }\href {\doibase 10.1098/rspa.1990.0055} {\bibfield  {journal}
  {\bibinfo  {journal} {Proc. Roy. Soc. London A.}\ }\textbf {\bibinfo {volume}
  {429}},\ \bibinfo {pages} {141} (\bibinfo {year} {1990})}\BibitemShut
  {NoStop}%
\bibitem [{\citenamefont {Paulsen}(2019)}]{Paulsen19}%
  \BibitemOpen
  \bibfield  {author} {\bibinfo {author} {\bibfnamefont {J.~D.}\ \bibnamefont
  {Paulsen}},\ }\href@noop {} {\bibfield  {journal} {\bibinfo  {journal}
  {Annual Review of Condensed Matter Physics}\ }\textbf {\bibinfo {volume}
  {10}},\ \bibinfo {pages} {431} (\bibinfo {year} {2019})}\BibitemShut
  {NoStop}%
\bibitem [{\citenamefont {Vella}(2019)}]{Vella19}%
  \BibitemOpen
  \bibfield  {author} {\bibinfo {author} {\bibfnamefont {D.}~\bibnamefont
  {Vella}},\ }\href {\doibase 10.1038/s42254-019-0063-1} {\bibfield  {journal}
  {\bibinfo  {journal} {{Nat. Rev. Physics}}\ }\textbf {\bibinfo {volume}
  {1}},\ \bibinfo {pages} {425} (\bibinfo {year} {2019})}\BibitemShut {NoStop}%
\bibitem [{\citenamefont {Brakke}(1992)}]{Brakke1992}%
  \BibitemOpen
  \bibfield  {author} {\bibinfo {author} {\bibfnamefont {K.~A.}\ \bibnamefont
  {Brakke}},\ }\href@noop {} {\bibfield  {journal} {\bibinfo  {journal}
  {Experimental Mathematics}\ }\textbf {\bibinfo {volume} {1}},\ \bibinfo
  {pages} {141} (\bibinfo {year} {1992})}\BibitemShut {NoStop}%
\bibitem [{\citenamefont {Nayyar}\ \emph
  {et~al.}(2011{\natexlab{b}})\citenamefont {Nayyar}, \citenamefont
  {Ravi-Chandar},\ and\ \citenamefont {Huang}}]{Nayya11}%
  \BibitemOpen
  \bibfield  {author} {\bibinfo {author} {\bibfnamefont {V.}~\bibnamefont
  {Nayyar}}, \bibinfo {author} {\bibfnamefont {K.}~\bibnamefont
  {Ravi-Chandar}}, \ and\ \bibinfo {author} {\bibfnamefont {R.}~\bibnamefont
  {Huang}},\ }\href
  {http://www.sciencedirect.com/science/article/pii/S0020768311003040}
  {\bibfield  {journal} {\bibinfo  {journal} {International Journal of Solids
  and Structures}\ }\textbf {\bibinfo {volume} {48}},\ \bibinfo {pages} {3471 }
  (\bibinfo {year} {2011}{\natexlab{b}})}\BibitemShut {NoStop}%
\bibitem [{\citenamefont {Healey}\ \emph {et~al.}(2013)\citenamefont {Healey},
  \citenamefont {Li},\ and\ \citenamefont {Cheng}}]{Healy13}%
  \BibitemOpen
  \bibfield  {author} {\bibinfo {author} {\bibfnamefont {T.~J.}\ \bibnamefont
  {Healey}}, \bibinfo {author} {\bibfnamefont {Q.}~\bibnamefont {Li}}, \ and\
  \bibinfo {author} {\bibfnamefont {R.-B.}\ \bibnamefont {Cheng}},\ }\href
  {\doibase 10.1007/s00332-013-9168-3} {\bibfield  {journal} {\bibinfo
  {journal} {J. Nonlin. Sci.}\ }\textbf {\bibinfo {volume} {23}},\ \bibinfo
  {pages} {777} (\bibinfo {year} {2013})}\BibitemShut {NoStop}%
\bibitem [{\citenamefont {Li}\ and\ \citenamefont {Healey}(2016)}]{Healy16}%
  \BibitemOpen
  \bibfield  {author} {\bibinfo {author} {\bibfnamefont {Q.}~\bibnamefont
  {Li}}\ and\ \bibinfo {author} {\bibfnamefont {T.~J.}\ \bibnamefont
  {Healey}},\ }\href {\doibase 10.1016/j.jmps.2015.12.001} {\bibfield
  {journal} {\bibinfo  {journal} {J. Mech. Phys. Solids}\ }\textbf {\bibinfo
  {volume} {97}},\ \bibinfo {pages} {260} (\bibinfo {year} {2016})},\ \bibinfo
  {note} {symposium on Length Scale in Solid Mechanics - Mathematical and
  Physical Aspects, Inst Henri Poincare, Paris, FRANCE, JUN 19-20,
  2014}\BibitemShut {NoStop}%
\bibitem [{\citenamefont {Sipos}\ and\ \citenamefont {Feher}(2016)}]{Sipos16}%
  \BibitemOpen
  \bibfield  {author} {\bibinfo {author} {\bibfnamefont {A.~A.}\ \bibnamefont
  {Sipos}}\ and\ \bibinfo {author} {\bibfnamefont {E.}~\bibnamefont {Feher}},\
  }\href {\doibase 10.1016/j.ijsolstr.2016.07.021} {\bibfield  {journal}
  {\bibinfo  {journal} {Int. J. Solids. Struct.}\ }\textbf {\bibinfo {volume}
  {97-98}},\ \bibinfo {pages} {275} (\bibinfo {year} {2016})}\BibitemShut
  {NoStop}%
\bibitem [{\citenamefont {Fu}\ \emph {et~al.}(2019)\citenamefont {Fu},
  \citenamefont {Wang}, \citenamefont {Xu}, \citenamefont {Huo},\ and\
  \citenamefont {Potier-Ferry}}]{Fu19}%
  \BibitemOpen
  \bibfield  {author} {\bibinfo {author} {\bibfnamefont {C.}~\bibnamefont
  {Fu}}, \bibinfo {author} {\bibfnamefont {T.}~\bibnamefont {Wang}}, \bibinfo
  {author} {\bibfnamefont {F.}~\bibnamefont {Xu}}, \bibinfo {author}
  {\bibfnamefont {Y.}~\bibnamefont {Huo}}, \ and\ \bibinfo {author}
  {\bibfnamefont {M.}~\bibnamefont {Potier-Ferry}},\ }\href {\doibase
  10.1016/j.jmps.2018.11.005} {\bibfield  {journal} {\bibinfo  {journal} {J.
  Mech. Phys. Solids}\ }\textbf {\bibinfo {volume} {124}},\ \bibinfo {pages}
  {446} (\bibinfo {year} {2019})}\BibitemShut {NoStop}%
\bibitem [{\citenamefont {Wang}\ \emph {et~al.}(2019)\citenamefont {Wang},
  \citenamefont {Fu}, \citenamefont {Xu}, \citenamefont {Huo},\ and\
  \citenamefont {Potier-Ferry}}]{Wang19}%
  \BibitemOpen
  \bibfield  {author} {\bibinfo {author} {\bibfnamefont {T.}~\bibnamefont
  {Wang}}, \bibinfo {author} {\bibfnamefont {C.}~\bibnamefont {Fu}}, \bibinfo
  {author} {\bibfnamefont {F.}~\bibnamefont {Xu}}, \bibinfo {author}
  {\bibfnamefont {Y.}~\bibnamefont {Huo}}, \ and\ \bibinfo {author}
  {\bibfnamefont {M.}~\bibnamefont {Potier-Ferry}},\ }\href {\doibase
  10.1016/j.ijengsci.2018.12.002} {\bibfield  {journal} {\bibinfo  {journal}
  {Int. J. Eng. Sci.}\ }\textbf {\bibinfo {volume} {136}},\ \bibinfo {pages}
  {1} (\bibinfo {year} {2019})}\BibitemShut {NoStop}%
\bibitem [{\citenamefont {Panaitescu}\ \emph {et~al.}(2019)\citenamefont
  {Panaitescu}, \citenamefont {Xin}, \citenamefont {Davidovitch}, \citenamefont
  {Chopin},\ and\ \citenamefont {Kudrolli}}]{Panaitescu2019}%
  \BibitemOpen
  \bibfield  {author} {\bibinfo {author} {\bibfnamefont {A.}~\bibnamefont
  {Panaitescu}}, \bibinfo {author} {\bibfnamefont {M.}~\bibnamefont {Xin}},
  \bibinfo {author} {\bibfnamefont {B.}~\bibnamefont {Davidovitch}}, \bibinfo
  {author} {\bibfnamefont {J.}~\bibnamefont {Chopin}}, \ and\ \bibinfo {author}
  {\bibfnamefont {A.}~\bibnamefont {Kudrolli}},\ }\href {\doibase
  10.1103/PhysRevE.100.053003} {\bibfield  {journal} {\bibinfo  {journal}
  {Phys. Rev. E}\ }\textbf {\bibinfo {volume} {100}},\ \bibinfo {pages}
  {053003} (\bibinfo {year} {2019})}\BibitemShut {NoStop}%
\bibitem [{\citenamefont {Landau}\ and\ \citenamefont
  {Lifshitz}(1986)}]{LandauTE}%
  \BibitemOpen
  \bibfield  {author} {\bibinfo {author} {\bibfnamefont {L.~D.}\ \bibnamefont
  {Landau}}\ and\ \bibinfo {author} {\bibfnamefont {E.}~\bibnamefont
  {Lifshitz}},\ }\href@noop {} {\emph {\bibinfo {title} {Theory of
  Elasticity}}}\ (\bibinfo  {publisher} {Pergamon},\ \bibinfo {address} {New
  York},\ \bibinfo {year} {1986})\BibitemShut {NoStop}%
\bibitem [{Note2()}]{Note2}%
  \BibitemOpen
  \bibinfo {note} {The error incurred by considering BCs through the
  undeformed, rather the deformed sheet, is a higher order in
  $T/Y$.}\BibitemShut {Stop}%
\bibitem [{\citenamefont {Benthem}(1963)}]{Benthem1963}%
  \BibitemOpen
  \bibfield  {author} {\bibinfo {author} {\bibfnamefont {J.~P.}\ \bibnamefont
  {Benthem}},\ }\href@noop {} {\bibfield  {journal} {\bibinfo  {journal} {The
  Quarterly Journal of Mechanics and Applied Mathematics}\ }\textbf {\bibinfo
  {volume} {16}},\ \bibinfo {pages} {413} (\bibinfo {year} {1963})}\BibitemShut
  {NoStop}%
\bibitem [{\citenamefont {Spence}(1983)}]{Spence83}%
  \BibitemOpen
  \bibfield  {author} {\bibinfo {author} {\bibfnamefont {D.~A.}\ \bibnamefont
  {Spence}},\ }\href@noop {} {\bibfield  {journal} {\bibinfo  {journal} {IMA.
  J. App. Math.}\ }\textbf {\bibinfo {volume} {30}},\ \bibinfo {pages} {107 }
  (\bibinfo {year} {1983})}\BibitemShut {NoStop}%
\bibitem [{\citenamefont {Chopin}\ \emph {et~al.}(2018)\citenamefont {Chopin},
  \citenamefont {Panaitescu},\ and\ \citenamefont {Kudrolli}}]{Chopin18}%
  \BibitemOpen
  \bibfield  {author} {\bibinfo {author} {\bibfnamefont {J.}~\bibnamefont
  {Chopin}}, \bibinfo {author} {\bibfnamefont {A.}~\bibnamefont {Panaitescu}},
  \ and\ \bibinfo {author} {\bibfnamefont {A.}~\bibnamefont {Kudrolli}},\
  }\href@noop {} {\bibfield  {journal} {\bibinfo  {journal} {Phys. Rev. E}\
  }\textbf {\bibinfo {volume} {98}},\ \bibinfo {pages} {043003} (\bibinfo
  {year} {2018})}\BibitemShut {NoStop}%
\bibitem [{\citenamefont {Moffatt}(1964)}]{Moffatt64}%
  \BibitemOpen
  \bibfield  {author} {\bibinfo {author} {\bibfnamefont {H.~K.}\ \bibnamefont
  {Moffatt}},\ }\href@noop {} {\bibfield  {journal} {\bibinfo  {journal} {J.
  Fluid. Mech.}\ }\textbf {\bibinfo {volume} {18}},\ \bibinfo {pages} {1 }
  (\bibinfo {year} {1964})}\BibitemShut {NoStop}%
\bibitem [{\citenamefont {Shankar}\ and\ \citenamefont
  {Deshpande}(2000)}]{Shankar2000}%
  \BibitemOpen
  \bibfield  {author} {\bibinfo {author} {\bibfnamefont {P.~N.}\ \bibnamefont
  {Shankar}}\ and\ \bibinfo {author} {\bibfnamefont {M.~D.}\ \bibnamefont
  {Deshpande}},\ }\href@noop {} {\bibfield  {journal} {\bibinfo  {journal}
  {Annual Review of Fluid Mechanics}\ }\textbf {\bibinfo {volume} {32}},\
  \bibinfo {pages} {93} (\bibinfo {year} {2000})}\BibitemShut {NoStop}%
\bibitem [{\citenamefont {Cerda}\ and\ \citenamefont
  {Mahadevan}(2003{\natexlab{b}})}]{Cerda-Mahadevan2003}%
  \BibitemOpen
  \bibfield  {author} {\bibinfo {author} {\bibfnamefont {E.}~\bibnamefont
  {Cerda}}\ and\ \bibinfo {author} {\bibfnamefont {L.}~\bibnamefont
  {Mahadevan}},\ }\href {\doibase 10.1103/PhysRevLett.90.074302} {\bibfield
  {journal} {\bibinfo  {journal} {Phys. Rev. Lett.}\ }\textbf {\bibinfo
  {volume} {90}},\ \bibinfo {pages} {074302} (\bibinfo {year}
  {2003}{\natexlab{b}})}\BibitemShut {NoStop}%
\end{thebibliography}%

\end{document}